\documentclass{ieeeconf}
\pdfoutput=1
{}
{}
{}

\usepackage{amsmath}
\usepackage{indentfirst}
\usepackage{booktabs,cases}
\usepackage{amssymb}
\usepackage{multirow,color}
\usepackage{mathrsfs}
\usepackage{bm}
\usepackage{graphicx}
\usepackage{float}
\usepackage{subfigure}
\usepackage{color}
\usepackage{url}
\newsavebox\CBox

\title{\LARGE \bf
	Color image segmentation based on a convex K-means approach}

\author{\authorblockN{Tingting Wu$^1$, Xiaoyu Gu$^1$, Jinbo Shao$^1$, Ruoxuan Zhou$^2$, Zhi Li$^{3,*}$}
	\thanks{$^*$The corresponding author of this paper.} 
	\thanks{$^1$School of Science, Nanjing University of Posts and Telecommunications, Nanjing, 210023, China.}
	\thanks{$^2$College of Oceanography and Space Informatics, China University of Petroleum, Qingdao, 266580, China.}
	\thanks{$^3$The Department of Computer Science and Technology, Shanghai Key Laboratory of Multidimensional Information Processing, East China Normal University, Shanghai, 200241, China. \textit{E-mail: zli@cs.ecnu.edu.cn.}}
}

\begin{document}
	
	\maketitle
	\thispagestyle{plain}
	\pagestyle{plain}
	\pagenumbering{arabic}
	\newcommand{\etal}{\textit{et al.}}
	\definecolor{ao}{rgb}{0.0, 0.5, 0.0}
	\makeatletter
	\def\algbackskip{\hskip-\ALG@thistlm}
	\makeatother
	
	\newcommand\mycommfont[1]{\footnotesize\ttfamily\textcolor{black}{#1}}

\begin{abstract}
Image segmentation is a fundamental and challenging task in image processing and computer vision. The color image segmentation is attracting more attention due to the color image provides more information than the gray image. In this paper, we propose a variational model based on a convex K-means approach to segment color images. The proposed variational method uses a combination of $l_1$ and $l_2$ regularizers to maintain edge information of objects in images while overcoming the staircase effect.
Meanwhile, our one-stage strategy is an improved version based on the smoothing and thresholding strategy, which contributes to improving the accuracy of segmentation. The proposed method performs the following steps.
First, we specify the color set which can be determined by human or the K-means method. Second, we use a variational model to obtain the most appropriate color for each pixel from the color set via convex relaxation and lifting. The Chambolle-Pock algorithm and simplex projection are applied to solve the variational model effectively. Experimental results and comparison analysis demonstrate the effectiveness and robustness of our method.
\end{abstract}

\maketitle

\section{Introduction}
\begin{figure*}[htb]
	\centering
	\subfigure[Given image]{
		\includegraphics[width=1.2in]{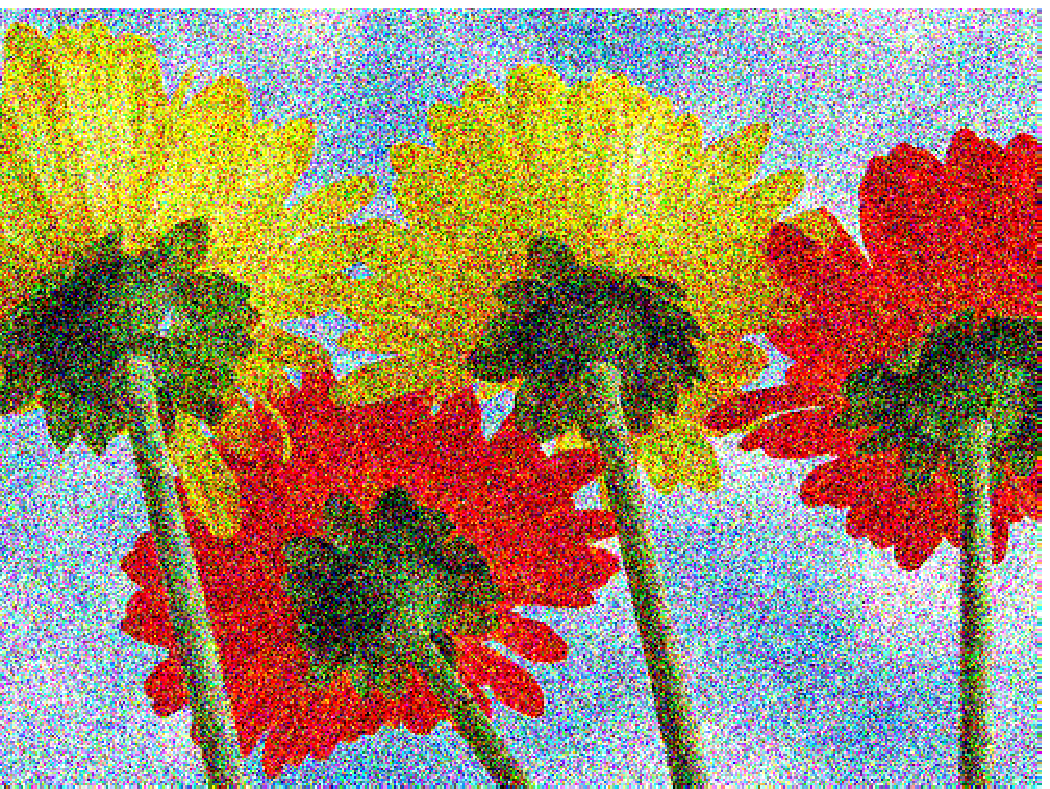}}
	\subfigure[SLaT \cite{cai2017three}]{
		\includegraphics[width=1.2in]{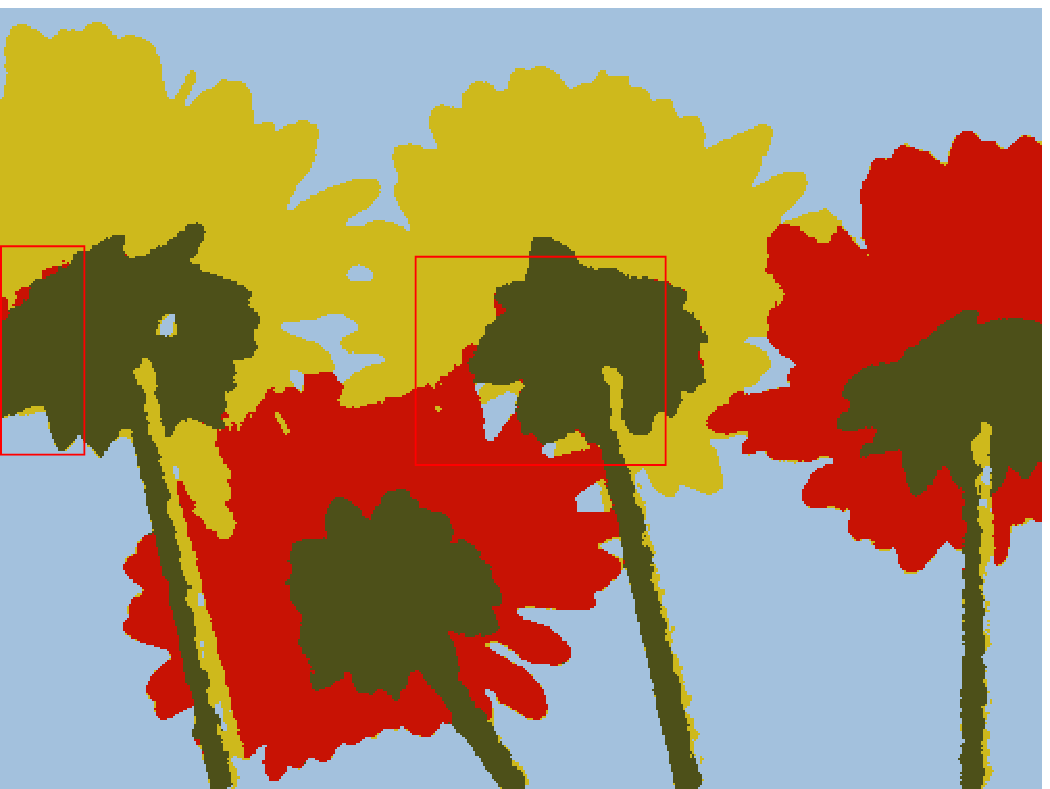}}
	\subfigure[Proposed]{
		\includegraphics[width=1.2in]{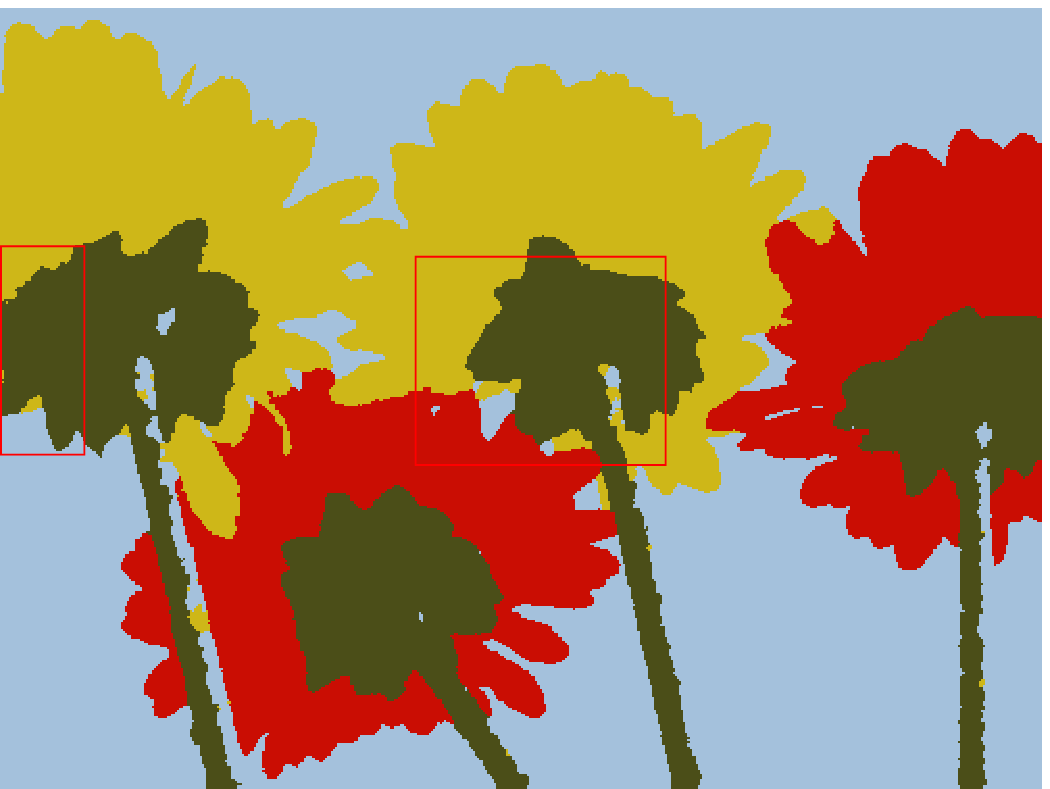}}
	\caption{Segmentation results of Flowers with Gaussian noise (mean $\tilde{\mu}=0$ and variance $\sigma^2=0.1$). Method \cite{cai2017three} used classical K-means framework (Figure (b)), where the pixels around the stalks are wrongly classified; our proposed method in Figure (c) achieves a relatively better classification, see the rectangle part.}
	\label{flowerstoshowconvexkmeans}
\end{figure*}
\begin{figure*}[htb]
	\centering
	\subfigure[Given image]{
		\includegraphics[width=1.2in]{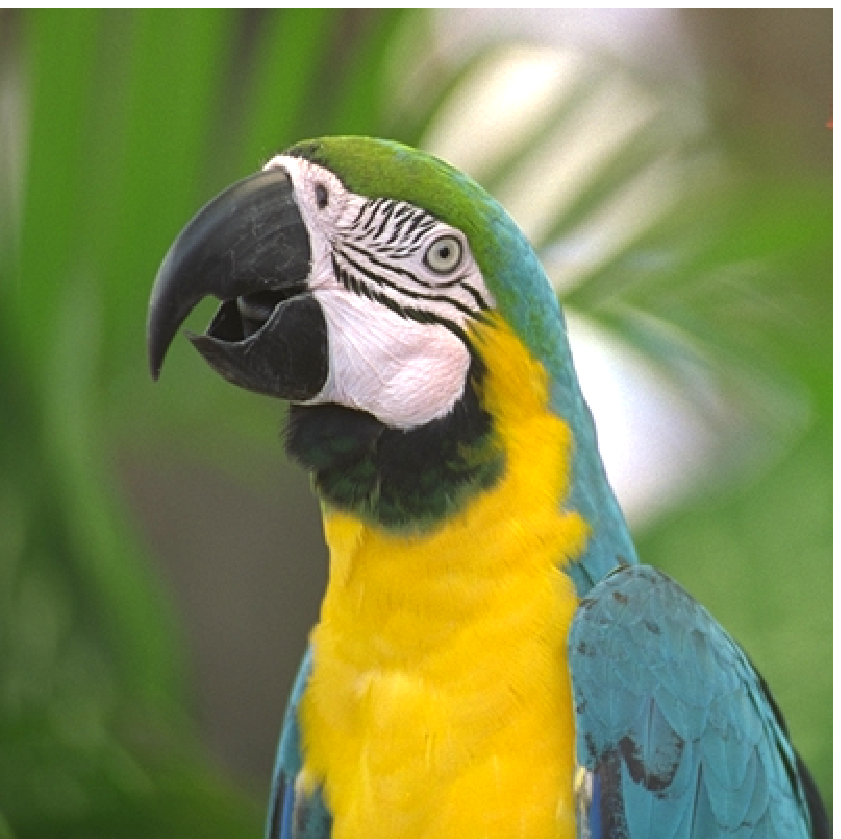}}	
	\subfigure[LC \cite{condat2017discrete} ]{
		\includegraphics[width=1.2in]{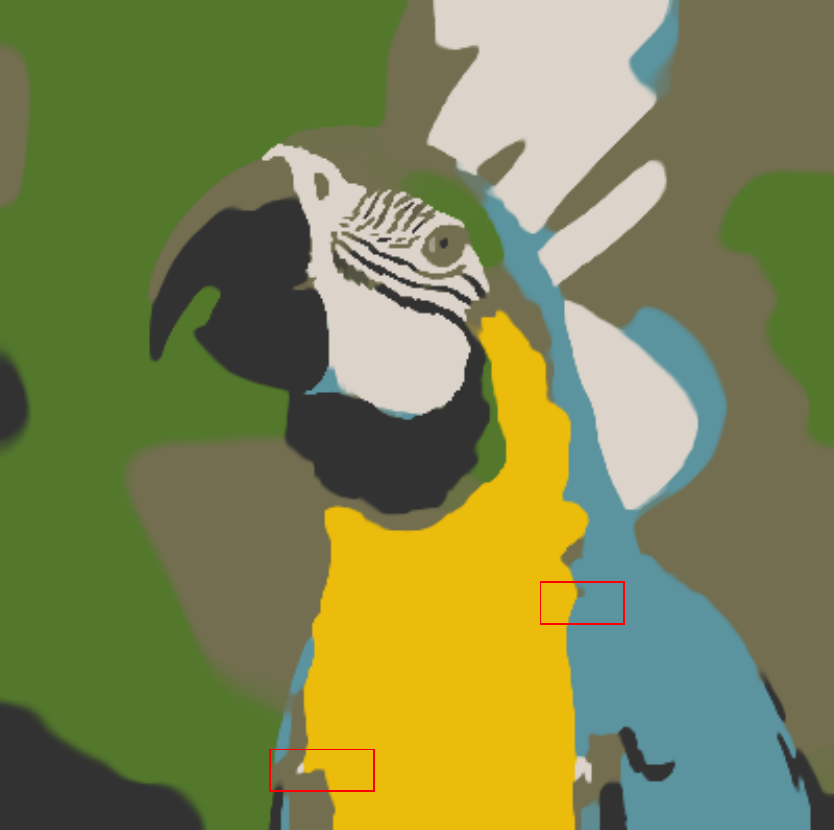}}
	\subfigure[Proposed]{
		\includegraphics[width=1.2in]{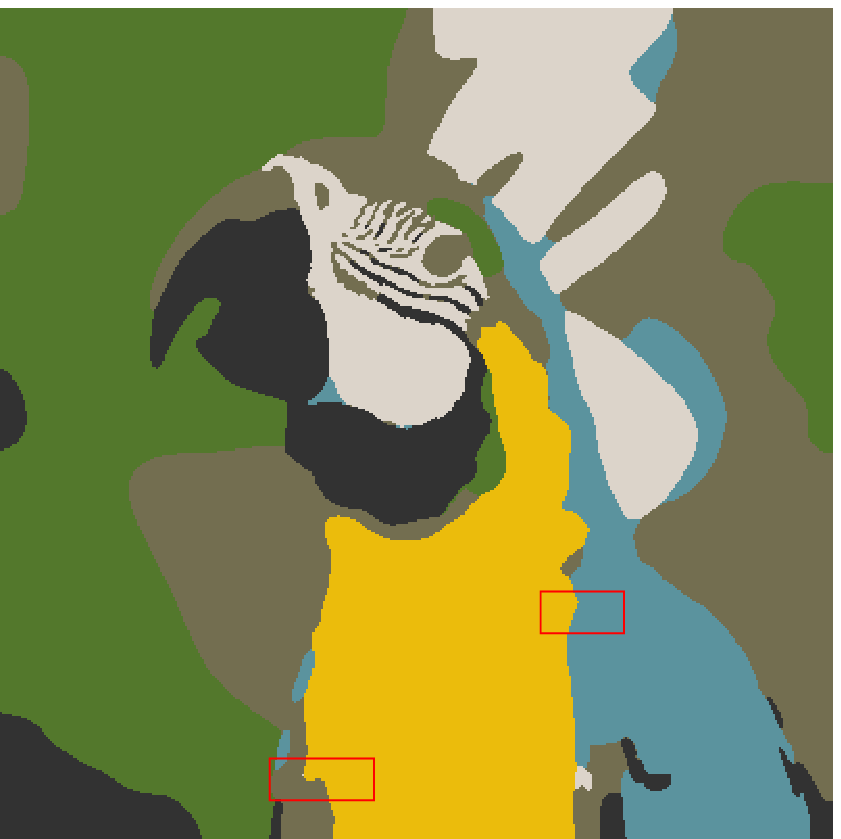}}
	\caption{Segmentation results of Parrot. Figure (b) produces some unwanted tiny structures \cite{condat2017discrete}; our result in Figure (c) can eliminate the unwanted tiny structures near the boundary by introducing the square of the first-order derivative term.}
	\label{parrottoshowsquare}
\end{figure*}

\label{intro}
Image segmentation is a fundamental and significant problem in image processing and computer vision, which has been widely applied in many fields such as object detection and medical diagnosis \cite{Crum2006Generalized,liu2017label,namburu2017generalised, mcconnon2012impact}. Image segmentation aims at dividing an image of $N$ pixels into $K$ regions with similar characteristics together (edges, intensities, colors or textures). 
Various models and algorithms have been extensively applied for image segmentation, including level-set methods \cite{liu2019binary,luo2019convex,wang2020level,yan2018convexity,rahmat2018comparison, wang2019the, wang2017efficient}, active contours \cite{chan2001active,ding2017active,jin2019active,wang2018active}, variational models \cite{mumford1989optimal, duan2014two, li2016two, cai2019linkage} and clustering methods \cite{zeng2017unified,ananthi2016new1, ananthi2016new2}, etc. 
Particularly, many successful methods for image segmentation are based on variational models and clustering methods. Variational models are flexible in structure, diverse in form, and superior in performance. The clustering methods are suitable and useful for both low- and high-dimensional data.

The Mumford-Shah (MS) model \cite{mumford1989optimal} is one of the most popular and influential variational segmentation models. The MS model proposed an energy minimization problem which approximates the true solution by finding optimal piecewise smooth approximations. However, it is very challenging to find or approximate its minimizer due to the MS model is nonconvex. It is worth noting that Cai \textit{et al.} proposed a two-stage strategy, \textit{i.e.}, smoothing and thresholding (SaT), for image segmentation \cite{cai2013two}. In the first stage, a convex optimization problem based on the MS model is solved to get a smooth approximation $u$ of the given image $f$. This stage could be regarded as the process of smoothing. Then in the second stage, they used the K-means algorithm or other clustering algorithms to obtain thresholds for segmentation, which can be seen as the process
of thresholding. The segmentation stage is independent of the optimization problem in the first stage, one can change the number of phases $K$ or thresholds $\rho$ without resolving the optimization problem.\par
Since the SaT strategy is a flexible methodology for image segmentation, it has been adopted and followed by many papers \cite{chan2018convex, duan2014two, li2016two, liu2018weighted, ma2018image, shi2016image, zhi2016two, Yijie2020A, cai2019linkage}. And the strategy was extended to images corrupted by Poisson or Gamma noises \cite{chan2014two}, color images \cite{cai2017three} and classification for high-dimensional data and point clouds. For example, literature \cite{cai2017three} proposed a three-stage method, \textit{i.e.}, Smoothing, Lifting and Thresholding, achieving a good result for segmenting images corrupted by noise. In \cite{cai2019linkage}, Cai \textit{et al.} revealed the connection between image restoration and segmentation by using the SaT strategy.
 
In the above segmentation methods based on the SaT strategy, thresholding is recognized as an effective method of image segmentation. A great number of thresholding techniques have been widely used for grayscale and color image segmentation. 	
Furthermore, some clustering algorithms have been used by many image segmentation models to determine thresholds for segmentation, such as the fuzzy clustering \cite{ananthi2016new1, ananthi2016new2, lei2018superpixel,namburu2017generalised, jia2020robust} and the K-means clustering \cite{cai2017three,cai2013two,chan2018convex,chan2014two,jin2013color,singh1996segmentation}, etc. In \cite{ananthi2016new1}, a thresholding method based on fuzzy set was proposed to deal with medical images, and the method got good performance visually and numerically. Jia \textit{et al.} \cite{jia2020robust} presented a robust self-sparse fuzzy clustering algorithm for solving over-segmentation.\par 
The K-means method is popular due to its simple principle and high calculation rate. 
In 2017, Condat proposed two methods for image segmentation based on a convex approach to the K-means method. The paper \cite{condat2017discrete} proposed a definition of discrete total variation (TV) and applied it to the segmentation of color images. Meanwhile, the method \cite{condat2017convex} showed a convex formulation of data clustering and image segmentation based on the ``Potts'' model (see Section \ref{Potts} for details), convex relaxation and lifting. However, the K centroids are pre-assigned by the author before the segmentation takes place. The above methods \cite{condat2017convex,condat2017discrete} introduced a large number of candidates which did not improve the image quality enough to be worth the extra computing cost, although avoided dealing with a non-convex problem. Furthermore, models in \cite{condat2017convex,condat2017discrete} are both based on TV regularization, which could cause the staircase effect, especially for noisy images.\par
 For the SaT strategy, the segmentation is carried in the thresholding stage and it uses the smoothed image obtained in the smoothing stage. The SaT strategy is quite important, and it can keep main features of the image while clear out the noise and meaningless features so that in all cases it can obtain good segmentation results. Besides, the K-means method has been widely used in segmentation method to determine the thresholds for the thresholding stage in the SaT strategy. However, Figure \ref{flowerstoshowconvexkmeans}(b) is the result of SLaT \cite{cai2017three} which used classical K-means framework to perform the segmentation. We can see that the pixels around the stalks are wrongly classified, there are many ``red dots'' around it (see the marked area). \par
 Thus the goals of this paper are summarized as follows:
 \begin{itemize}
 	\item [a.] to avoid solving non-convex and combinatorial problem for the sake of stability;
 	\item [b.] to reduce the staircase effect in the model based on the TV regularization;
 	\item [c.] to inherit the advantages of the SaT strategy especially the smoothing stage;
 	\item [d.] to adopt a new strategy for segmentation to determine the most appropriate color for each pixel.\end{itemize}\par

In summary, we propose a new method to achieve the above goals. The proposed method combines convex relaxation based on the K-means method and lifting. We perform an improved convex relaxation for the classical K-means method and transform the problem into an assignment problem by the technique of lifting \cite{cai2017three,chambolle2012convex,condat2017convex,condat2017discrete,pock2010global}. On the one hand, we add a square of the first-order derivative term based on the variational method \cite{condat2017convex} to reduce the staircase effect. The segmentation performance can be well guaranteed by the combination of $l_1$ and $l_2$ regularizers. On the other hand, we use the strategy of lifting to determine the most appropriate color for each pixel from the color set, which could overcome the drawback of the classical K-means method in color image segmentation.

The organization of this paper is as follows. In Section 2, we give the basic knowledge used throughout the paper. In Section 3, we propose our model for image segmentation and present the Chambolle-Pock algorithm to solve the suggested model. In Section 4, we present numerical experiments and some comparisons to show the superiority of our one-stage method. Conclusions are drawn in Section 5.\par

\begin{figure*}[htb]
	\centering
	\subfigure[Three-phases]{
		\includegraphics[width=1.1in]{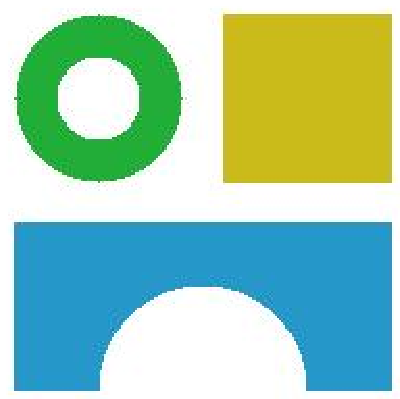}}
	\subfigure[Palette $M=4$]{
		\includegraphics[width=1.1in]{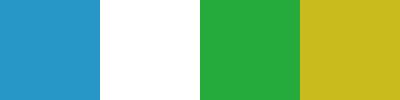}}	
	\subfigure[Six-phases]{
		\includegraphics[width=1.1in]{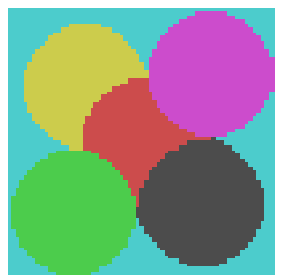}}
	\subfigure[Palette $M=6$]{
		\includegraphics[width=1.1in]{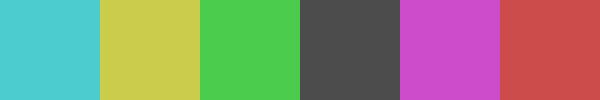}}
	\caption{The original synthetic images and corresponding Palettes with Three-phases and Six-phases, respectively.}
	\label{testimages1}
\end{figure*}

\begin{figure*}[htb]
	\centering
	\subfigure[Airplane]{
		\includegraphics[width=1in,height=0.8in]{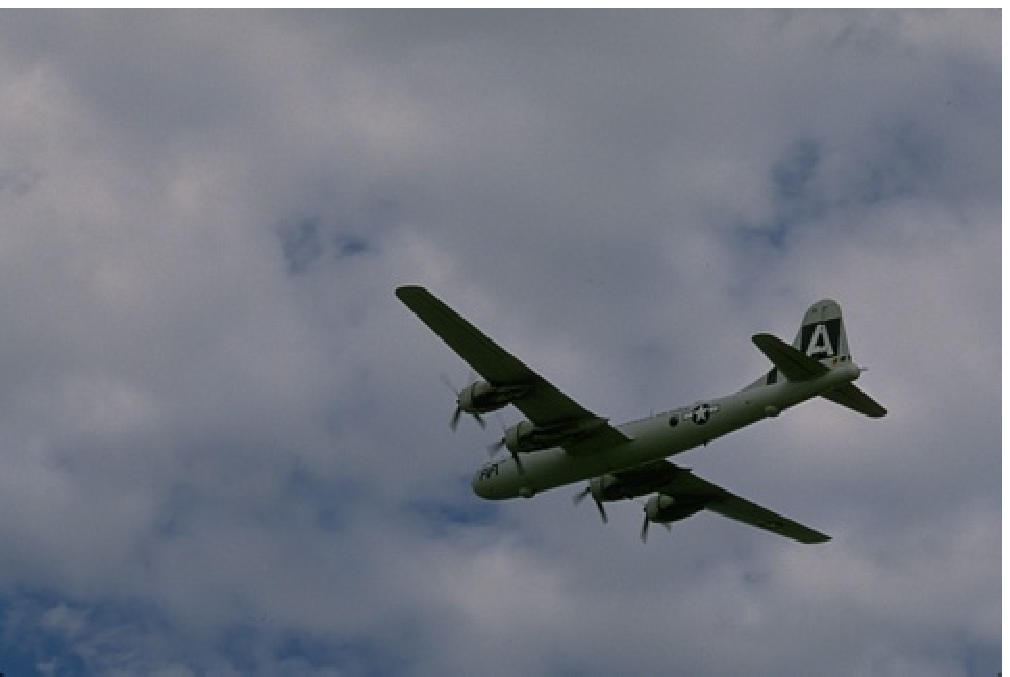}}
	\subfigure[Hill]{
		\includegraphics[width=1in,height=0.8in]{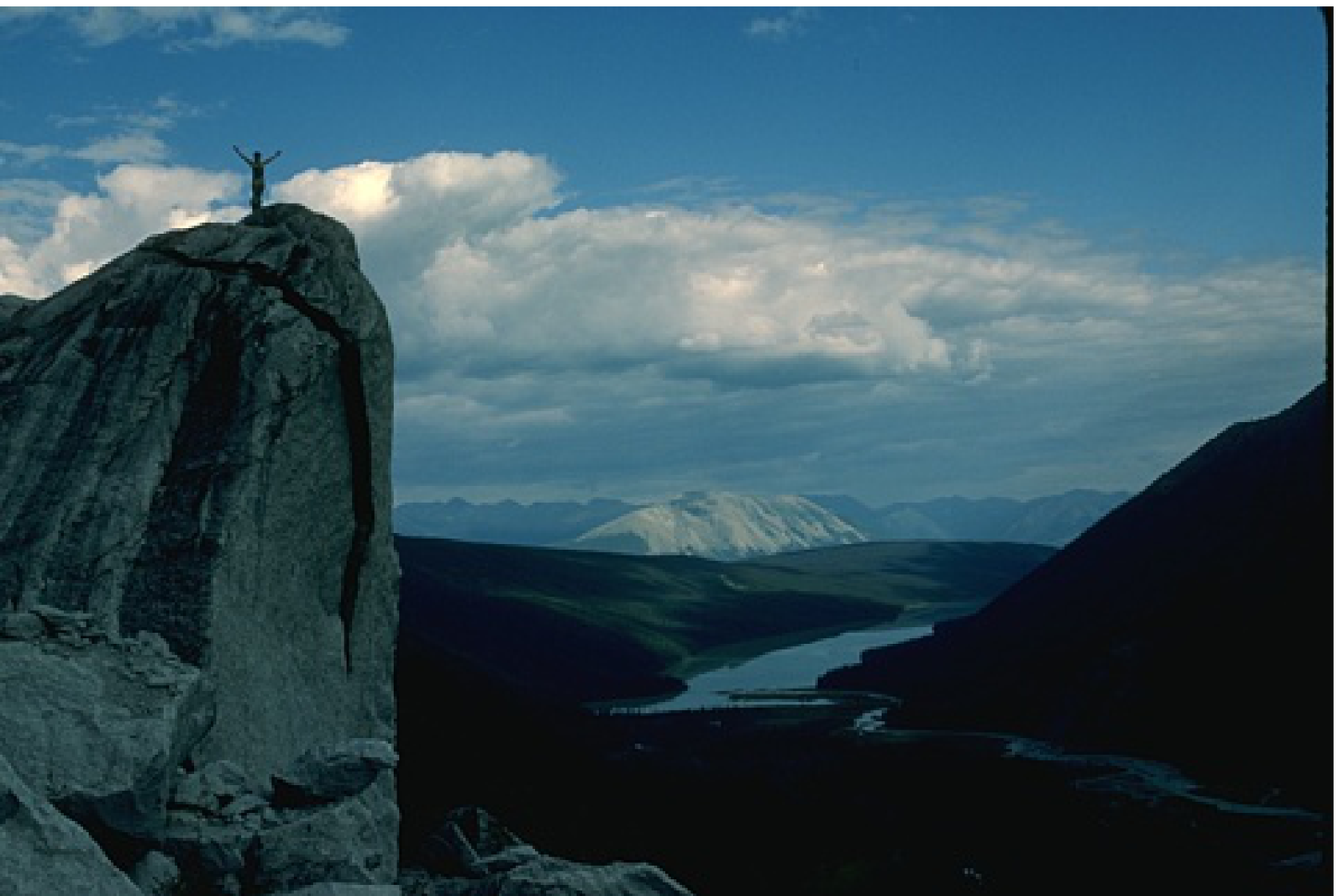}}
	\subfigure[Flowers]{
		\includegraphics[width=1in,height=0.8in]{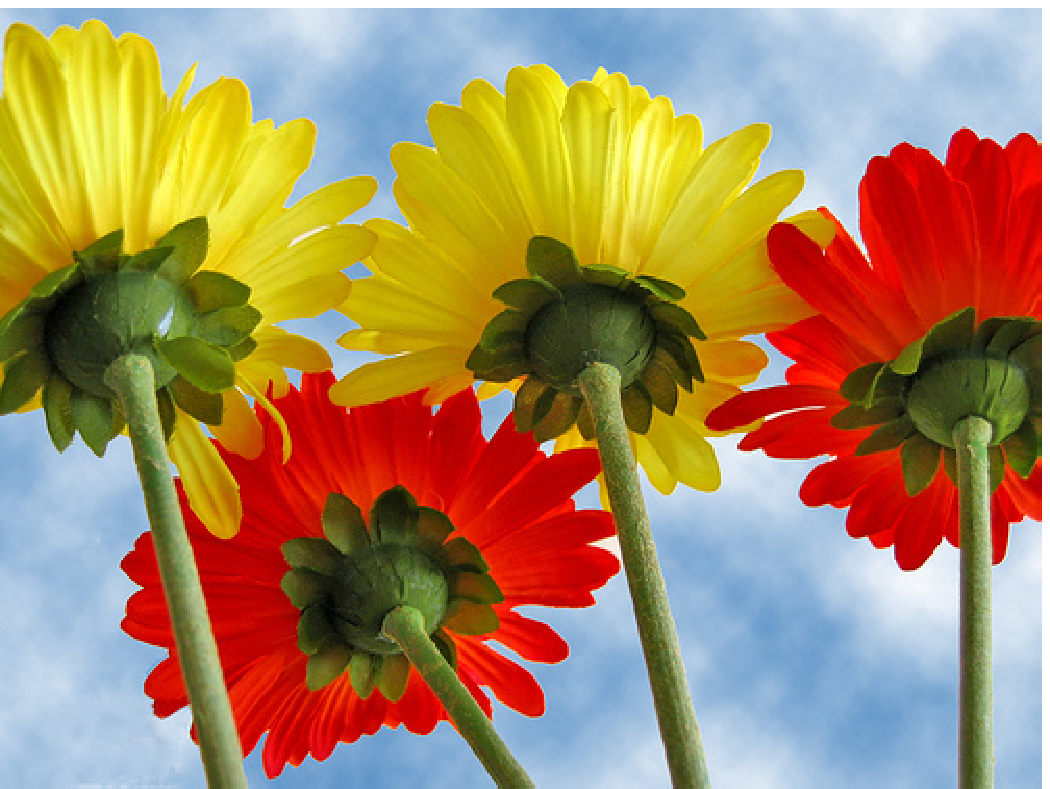}}
	\subfigure[Parrot]{
		\includegraphics[width=1in,height=0.8in]{clean/parrot/clean.eps}}
	\subfigure[Ladybug ]{
		\includegraphics[width=1in,height=0.8in]{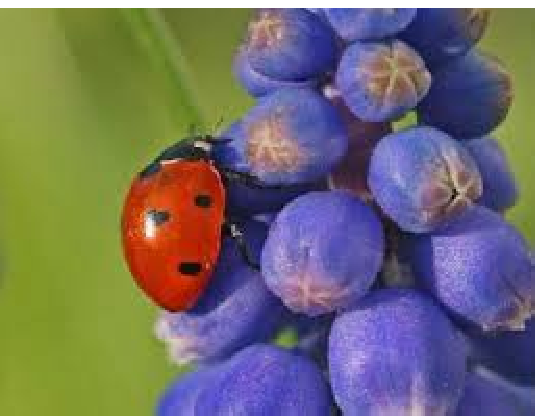}}
	\subfigure[Sunflowers]{
		\includegraphics[width=1in,height=0.8in]{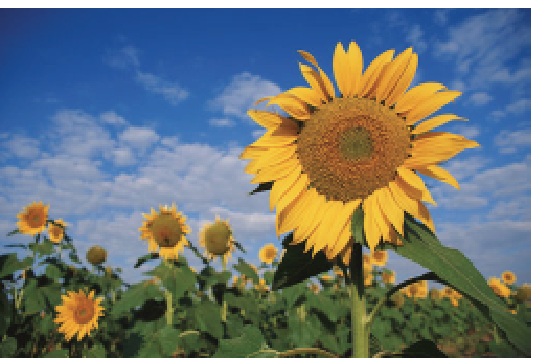}}
	\caption{The original real-world images with different color sets and structures.}
	\label{testimages2}
\end{figure*}
\begin{figure*}[htb]
	\centering	
		\subfigure[Palette $M=2$]{
		\includegraphics[width=1in]{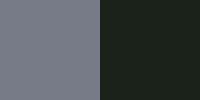}}
	\subfigure[Palette $M=3$]{
		\includegraphics[width=1in]{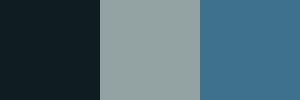}}
	\subfigure[Palette $M=4$]{
		\includegraphics[width=1in]{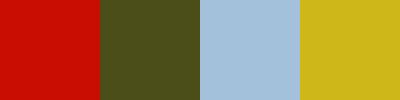}}
	\subfigure[Palette $M=6$]{
		\includegraphics[width=1in]{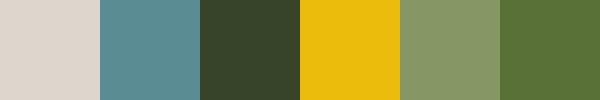}}
	\subfigure[Palette $M=6$]{
		\includegraphics[width=1in]{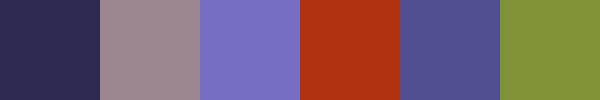}}
	\subfigure[Palette $M=6$]{
		\includegraphics[width=1in]{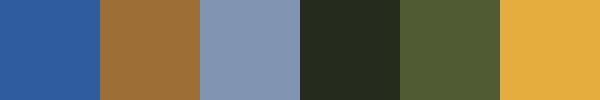}}	
	\caption{The used palettes in Figure \ref{testimages2}(a)-(f).}
	\label{cleansixphases_p}
\end{figure*}

\section{Model}
\subsection{Potts model}\label{Potts}
Firstly, we give some definitions for the classical K-means method in image processing. The data $f=(f_{n})_{n \in \Omega}$ is a 2D image  with values $f_n$, where domain $\Omega=\{1,\cdot\cdot\cdot,N_{1}\}\times\{1,\cdot\cdot\cdot,N_{2}\}$, and $n=N_{1} \times N_{2}$. The classical K-means problem minimizes the sum of Euclidean distance from every points $f_{n}$ to the nearest centroid $c_{k}$, aiming at partitioning $N$ points of $\mathcal{R}^{d}$ into $K$ groups, \textit{i.e.,} it solves the following problem to partition points into different clusters
\begin{equation}\label{pottsmodel1}
	\underset{(\Omega_{k})_{k=1}^{K}, (c_{k})_{k=1}^{K}}{\operatorname{minimize}} \frac{1}{2} \sum_{k=1}^{K} \sum_{n \in \Omega_{k}}\left\|f_{n}-c_{k}\right\|^{2}.
\end{equation}\par
Secondly, we consider a general problem based on the classical K-means problem.
Given an integer $K \geq 2$, one wants to partition
$\Omega$ into $K$ regions $\Omega_{k}$ (so $\bigcup^{K}_{k=1} \Omega_{k}=\Omega$ and $\Omega_{k^{'}}\bigcap \Omega_{k}=\emptyset$, for all $k\neq k^{'}$),
and to find the corresponding centroids $c_{k} \in \mathcal{R}^{d}$, so as to
\begin{equation}\label{pottsmodel}
	\underset{(\Omega_{k})_{k=1}^{K}, (c_{k})_{k=1}^{K}}{\operatorname{minimize}} \frac{1}{2} \sum_{k=1}^{K} \sum_{n \in \Omega_{k}}\left\|f_{n}-c_{k}\right\|^{2}+\frac{\lambda}{2} \sum_{k=1}^{K} \operatorname{per}\left(\Omega_{k}\right).
\end{equation}\par
In the above model (\ref{pottsmodel}), the first term is the data fidelity term, the second term is the penalization of the sum of the region perimeter (the region perimeter is $\operatorname{per}\left(\Omega_{k}\right)$, and $\operatorname{per}$ denotes the perimeter). The parameter $\lambda$ is to control the level of spatial regularization, which has been adopted in the NP-hard piecewise constant Mumford-Shah problem \cite{bar2014mumford,mumford1989optimal} to obtain spatial homogeneity by penalizing the sum of the region perimeter. The model is named as ``Potts'' (Penalization of the region perimeter) model. The classical K-means problem is the special case of the equation (\ref{pottsmodel}), then we can partition the point cloud $f=(f_{n})_{n \in \Omega}$ whose $c_{k}$ are the means. Please see \cite{condat2017convex,condat2017discrete} for more details.

\subsection{Hill-climbing method}
Traditional hill-climbing segmentation \cite{ding2011fast,ohashi2003hill} is a simple, fast algorithm that clusters colors of an image without any hand-tuning of parameters. Li \textit{et al.} made a few modifications to get a more stable hill-climbing procedure \cite{li2016two}. Moreover, the most significant advantage of hill-climbing is that it does not need prior knowledge of the number of clusters or the content of the given image and can detect the number of segments. As papers \cite{chandana2014clustering,li2016two} mentioned, the number of peaks is equal to the number of clusters, \textit{i.e.,} the number of segments ($K$). This process is quite simple to implement: we just group the matrix $f$ into a 3D histogram, then the hill-climbing algorithm attempts to locate the peaks on the color histogram \cite{li2016two}. For more details, see \cite{chandana2014clustering,ding2011fast,li2016two,ohashi2003hill} and references therein.

\subsection{The smoothed Potts model}\label{The smoothed Potts model}
As discussed in the above subsection, we first use the hill-climbing procedure to detect the segments ($K$) in advance and then we use the K-means algorithm to get the set $\Sigma={c_{k} \in[0,1]^{3}}$, where $k=1,\ldots,K$  of
$K\geq 2$ colors $c_k$, denoting the triplets of R, G, B values. We can use the {\sc Matlab} K-means function ``kmeans'' to get the $c_{k}$. Here, it is worth mentioning that there is no doubt that the color set could be specified by human or the K-means method. We can choose the most suitable color for each pixel by the segmentation accuracy to get a finer result.\par
We can define the segmented RGB color image $u$, whose R, G, B values at every pixel is one of the $c_{k}$, close to $f$ (also known as piecewise constant MS problem).\par
In the segmented image $u$, every element $u_{n}$ is one of the centroids
$c_{k}$, thus we can use the lifting strategy \cite{cai2017three,chambolle2012convex,condat2017convex,condat2017discrete,pock2010global} to reformulate the problem (\ref{pottsmodel}) as an equivalent assignment problem, then the unknown variable is the assignment array $z$  indexed by $k=1, \ldots, K$ with one more dimension than $f$. This reformulation meets the goal (d): to adopt a new strategy for segmentation to determine the most appropriate color for each pixel.\par
The assignment array $z$ has the following properties:
\begin{itemize} 
	\item [a.] The assignment array $z$  belongs to the set $\mathcal{A}$ of binary assignment vectors.
	\item [b.] For every $n \in \Omega$ and $k=1, \ldots, K$, if $u_{n}=c_{k}$ then $z_{k, n} \in \mathcal{A}^{\Omega}$ is equal to 1 and to 0 else, \textit{i.e.,} we want to choose a $c_{k}$ from the color set to represent the pixel of the $n \in \Omega$.
	\item [c.] Each vector $z_{ :, n}=\left(z_{k, n}\right)_{k=1}^{K}$ with elements in ${0,1}$ and its sum is 1.
\end{itemize}
Based on the above properties, we can get $u$ by a simple summation
\begin{equation}\label{segmentedu}
	u_{n}=\sum_{k=1}^{K} z_{k, n} c_{k}, \quad \forall n \in \Omega.
\end{equation}
Then, in short $\langle z, w\rangle=\frac{1}{2}\left\|f-u\right\|^{2}$, where $w_{k, n}=\frac{1}{2}\left\|f_{n}-c_{k}\right\|^{2}$.\par

The TV of the indicator function of set (1 inside, 0 outside) is equal to the perimeter of that set \cite{chambolle2012convex}. We can rewrite the second term in (\ref{pottsmodel}) as
\begin{align}\label{discretetv}
	\frac{\lambda}{2} \sum_{k=1}^{K} \operatorname{per}\left(\Omega_{k}\right)=\frac{\lambda}{2} \|\nabla z\|_{1}.
	\end{align}
Here we have some discrete form of the TV, and $\|\nabla z\|_{1}$ denotes the discrete total variation of $z$, \textit{i.e.,}
$\|\nabla z\|_{1}=\sum_{k} \sqrt{\left(\nabla_{x} z\right)_{k}^{2}+\left(\nabla_{y} z\right)_{k}^{2}}$ with $\nabla_{x}$ and $\nabla_{y}$ corresponding to the discrete derivative operators in the $x$-direction and $y$-direction, respectively.\par
As we know, image processing models based on TV regularization may preserve important edge information, but the staircase effect may be introduced. To reduce the staircase effect \cite{cai2013two}, we add a square of the first-order derivative in (\ref{discretetv}). In short,
we denote
\begin{equation}\label{smoothtv}
	\lambda \sum_{k=1}^{K} \operatorname{per}\left(\Omega_{k}\right):=\lambda\|\nabla z\|_{1}+\frac{\mu}{2}\|\nabla z\|_{2}^{2}.
\end{equation}
The effect of the square of the first-order derivative term can be seen in Figure \ref{parrottoshowsquare}. 

As we can see, when we set $\mu=0$, model \ref{smoothtv} becomes the model of \cite{condat2017discrete}. Comparing Figure \ref{parrottoshowsquare}(b) with (c), we can find the
(b) has some unwanted tiny structures (see the marked area). We add the square of the first-order derivative term, which is a smooth term.
This term works and smoothes out the individual dots, see Figure \ref{parrottoshowsquare}(c). More precisely, Figure \ref{parrottoshowsquare}(b) produces unwanted tiny structures as the method of \cite{condat2017discrete} based on TV regularization; Figure \ref{parrottoshowsquare}(c) can eliminate the unwanted tiny structures near the boundary by introducing the square of the first-order derivative term. So, it is of great significance to add this smooth term.  {This modification meets the goal (b) and (c): to reduce the staircase effect and inherit the advantages of the smoothing stage.}
\begin{figure*}[htb]
	\centering
	\subfigure[$\tilde{\mu}=0$, $\sigma^2=0.1$ ]{
		\includegraphics[width=1in]{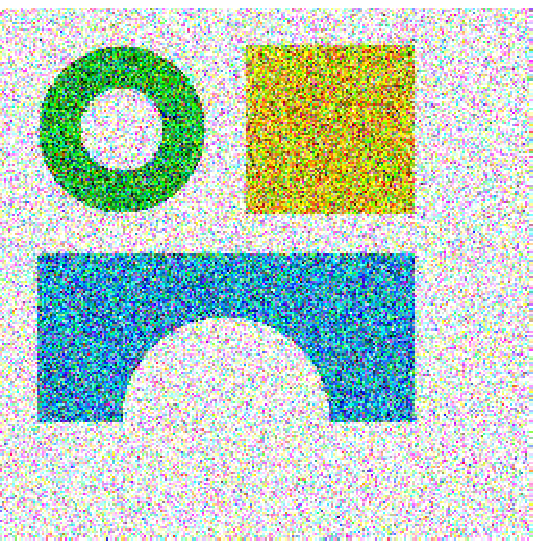}}
	\subfigure[$\tilde{\mu}=0$, $\sigma^2=0.3$ ]{
		\includegraphics[width=1in]{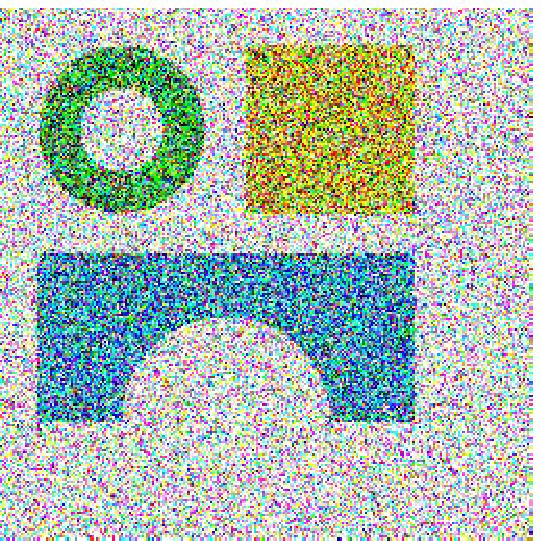}}
	\subfigure[$\tilde{\mu}=0$, $\sigma^2=0.5$ ]{
		\includegraphics[width=1in]{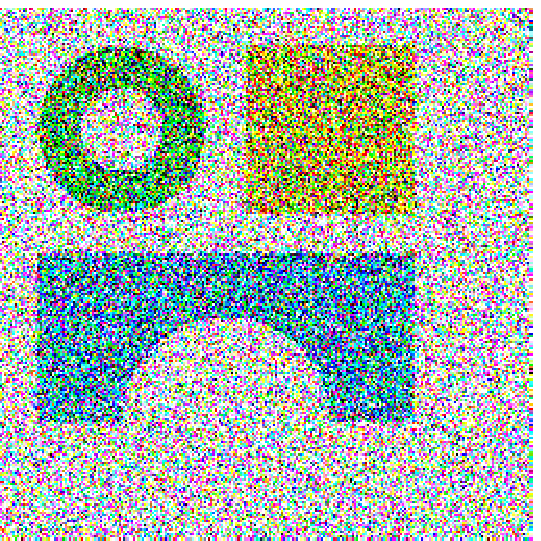}}
	\subfigure[$\tilde{\mu}=0$, $\sigma^2=0.1$ ]{
		\includegraphics[width=1.18in]{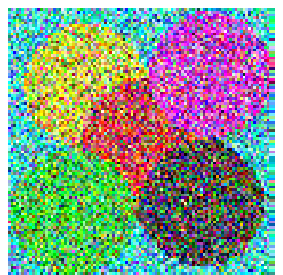}}
	\caption{The degraded images for Figure \ref{testimages1}(a) and (c). For Figure (a)-(c), we add the Gaussian noise with mean $\tilde{\mu}=0$ and variance $\sigma^2=0.1$, $\sigma^2=0.3$, and $\sigma^2=0.5$, respectively. For Figure (d), we add the Gaussian noise with mean $\tilde{\mu}=0$ and variance $\sigma^2=0.1$.}
	\label{noise_images}
\end{figure*}

\begin{figure*}[htb]
	\centering
	\subfigure[FRC \cite{li2010multiphase}]{	
		\includegraphics[width=1.in]{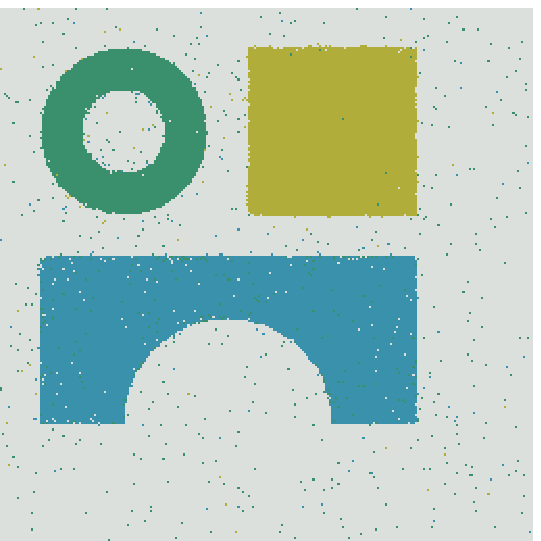}}
	\subfigure[DPP \cite{storath2014fast}]{	
		\includegraphics[width=1.in]{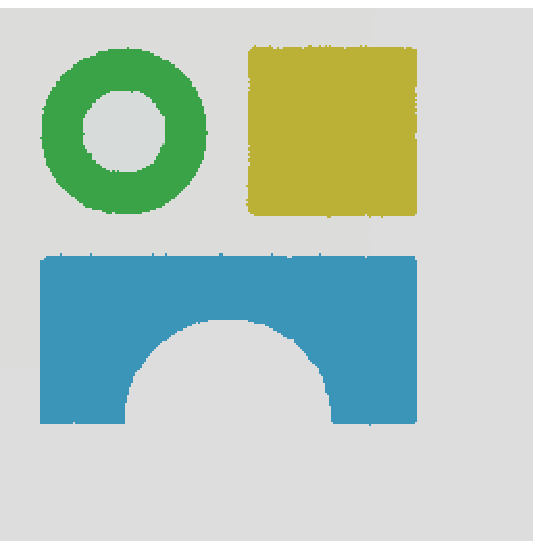}}
	\subfigure[SLaT \cite{cai2017three}]{	
		\includegraphics[width=1.in]{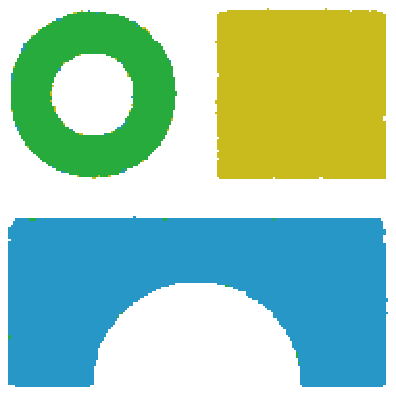}}
	\subfigure[LC \cite{condat2017discrete}]{	
		\includegraphics[width=1.in]{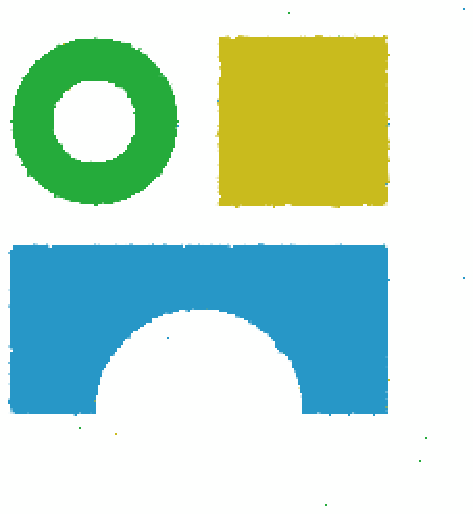}}
	\subfigure[RSSFC \cite{jia2020robust}]{	
		\includegraphics[width=1.in]{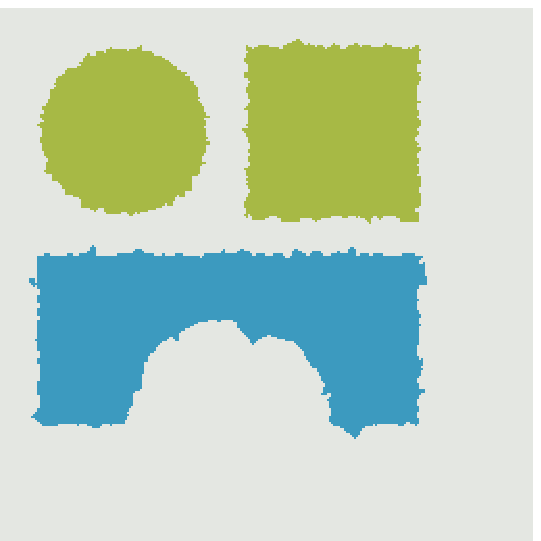}}
	\subfigure[Proposed]{	
		\includegraphics[width=1.in]{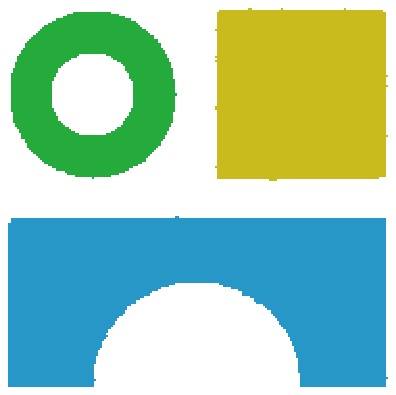}}
	\caption{The segmentation results of Three-shapes in Figure \ref{noise_images}(a). We add the Gaussian noise with mean $\tilde{\mu}=0$, variance $\sigma^2=0.1$.}
	\label{cleansixphases1}
\end{figure*}

\begin{figure*}[htb]
	\centering
	\subfigure[FRC \cite{li2010multiphase}]{	
		\includegraphics[width=1in]{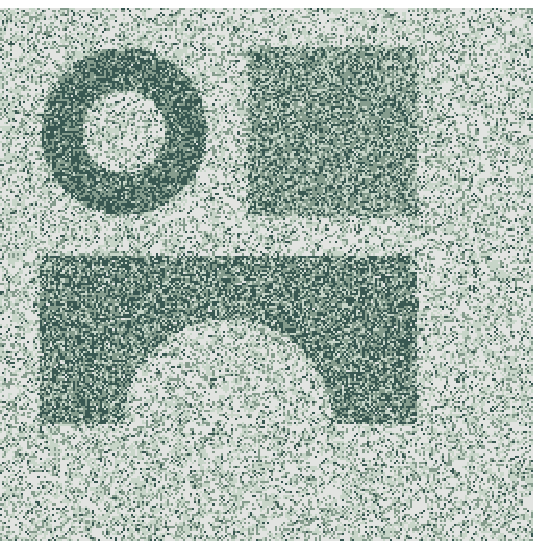}}
	\subfigure[DPP \cite{storath2014fast}]{	
		\includegraphics[width=1in]{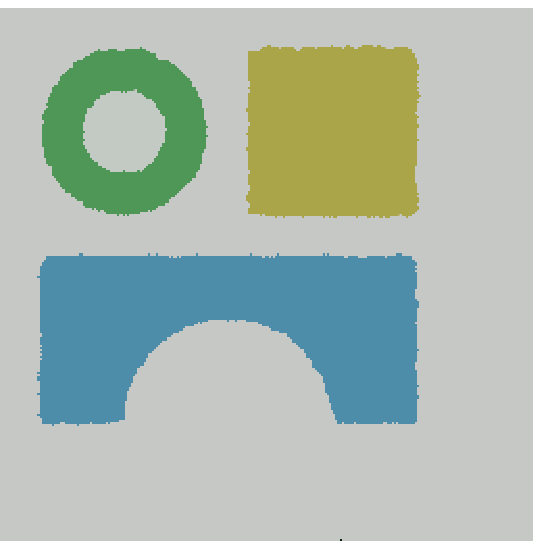}}
	\subfigure[SLaT \cite{cai2017three}]{	
		\includegraphics[width=1in]{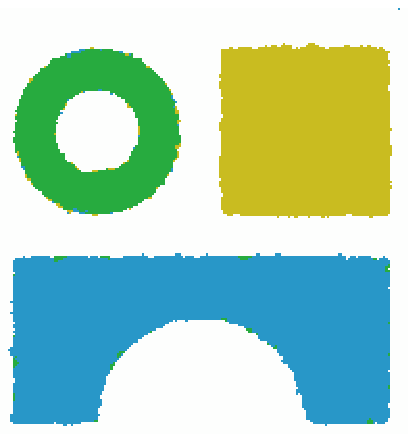}}
	\subfigure[LC \cite{condat2017discrete}]{	
		\includegraphics[width=1in]{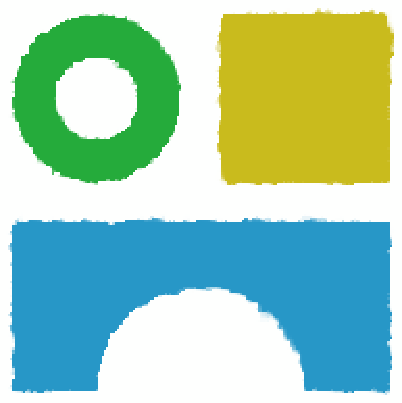}}
	\subfigure[RSSFC \cite{jia2020robust}]{	
		\includegraphics[width=1in]{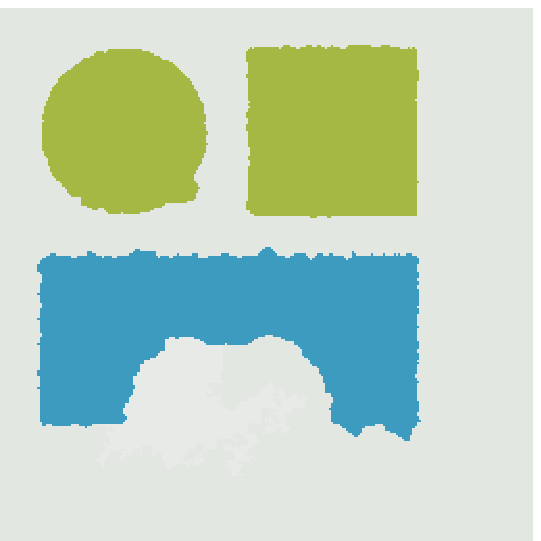}}
	\subfigure[Proposed]{	
		\includegraphics[width=1in]{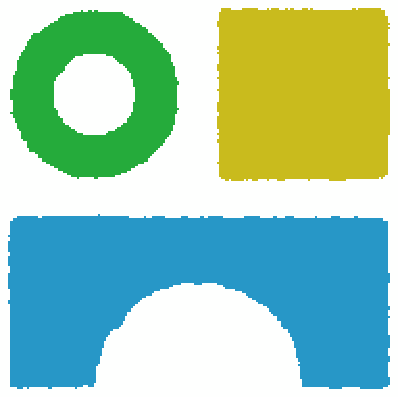}}
	\caption{The segmentation results of Three-shapes in Figure \ref{noise_images}(b). We add the Gaussian noise with mean $\tilde{\mu}=0$, variance $\sigma^2=0.3$.}
	\label{cleansixphases2}
\end{figure*}
\begin{figure*}[htb]
	\centering
	\subfigure[FRC \cite{li2010multiphase}]{	
		\includegraphics[width=1.in]{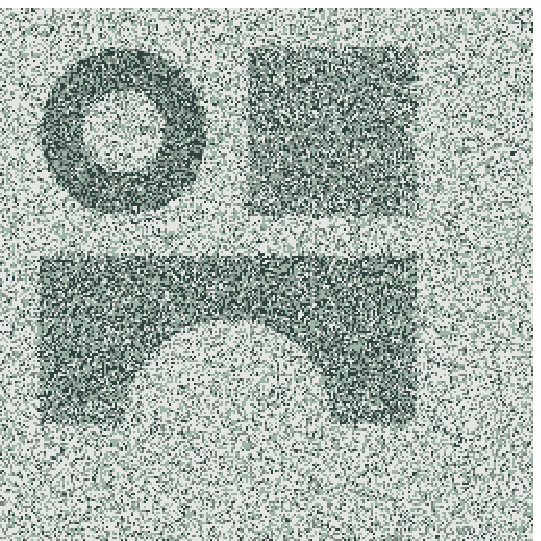}}
	\subfigure[DPP \cite{storath2014fast}]{	
		\includegraphics[width=1.in]{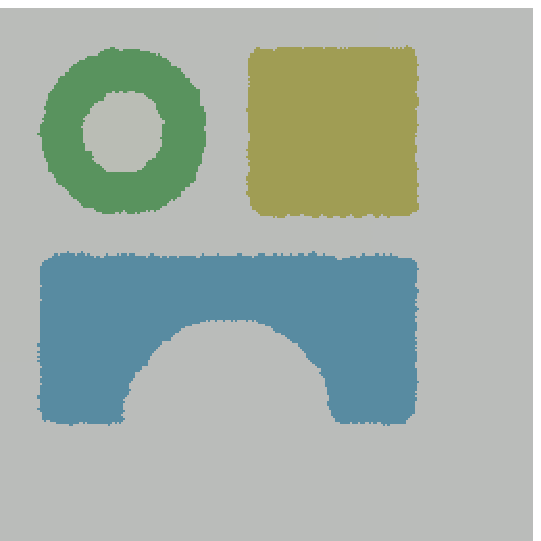}}
	\subfigure[SLaT \cite{cai2017three}]{	
		\includegraphics[width=1.in]{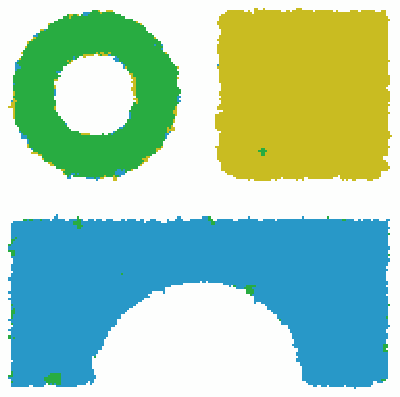}}
	\subfigure[LC \cite{condat2017discrete}]{	
		\includegraphics[width=1.in]{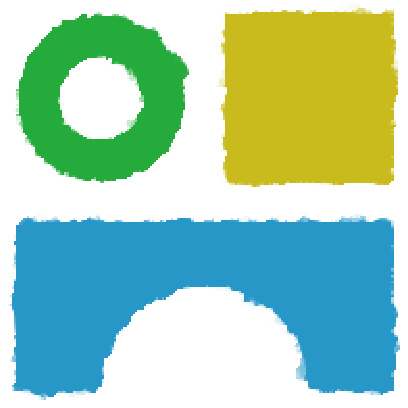}}
	\subfigure[RSSFC \cite{jia2020robust}]{	
		\includegraphics[width=1.in]{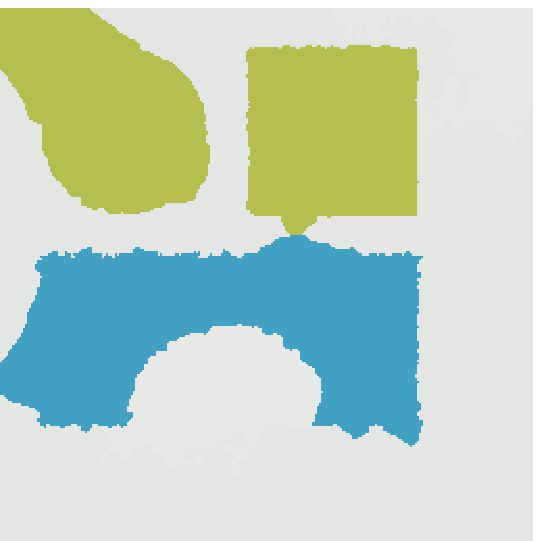}}
	\subfigure[Proposed]{	
		\includegraphics[width=1.in]{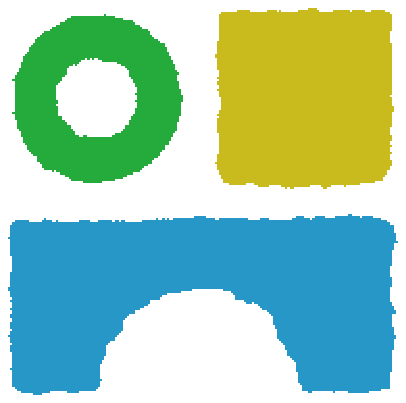}}
	\caption{The segmentation results of Three-shapes in Figure \ref{noise_images}(c). We add the Gaussian noise with mean $\tilde{\mu}=0$, variance $\sigma^{2}=0.5$.}
	\label{cleansixphases3}
\end{figure*}

\begin{table}[htb]
	\caption{The runtime in seconds of the color set by the K-means method and human specified (including the time taken in deciding the number of colors).}
	\centering
	\setlength{\tabcolsep}{5mm}{
		\scriptsize
		\begin{tabular}{c|c|c}
			\hline
			Images     & K-means & Human   \\ \hline
			Figure \ref{noise_images}(a)     & 8.36   & 157.67 \\
			Figure \ref{noise_images}(b)     & 8.36   & 159.54 \\
			Figure \ref{noise_images}(c)     & 8.39   & 159.93 \\ 
			Figure \ref{noise_images}(d)     & 6.18   & 275.27 \\\hline
	\end{tabular}}\label{Table4}
\end{table}
\begin{figure*}[htb]
	\centering
	\subfigure[FRC \cite{li2010multiphase}]{
		\includegraphics[width=1.in]{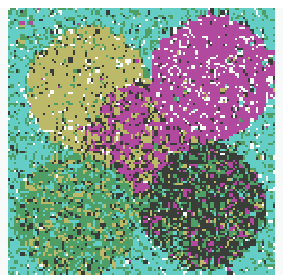}}
	\subfigure[DPP \cite{storath2014fast}]{
		\includegraphics[width=1.in]{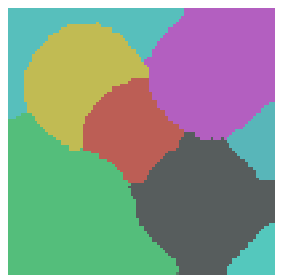}}
	\subfigure[SLaT \cite{cai2017three}]{
		\includegraphics[width=1.in]{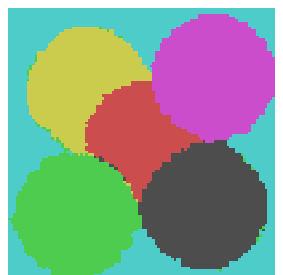}}
	\subfigure[LC \cite{condat2017discrete}]{
		\includegraphics[width=1in]{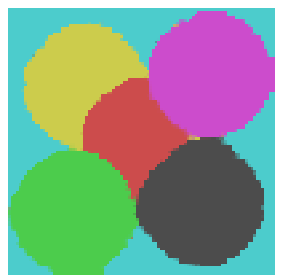}}
	\subfigure[RSSFC \cite{jia2020robust}]{	
		\includegraphics[width=1.in]{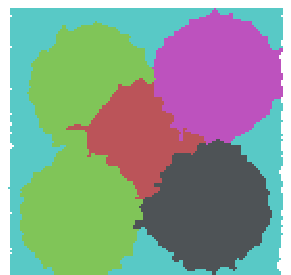}}
	\subfigure[Proposed]{
		\includegraphics[width=1.in]{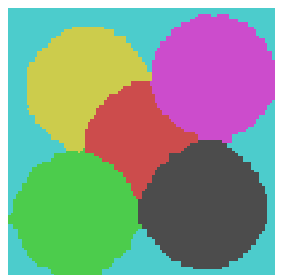}}
	\caption{The segmentation results of Six-phases in Figure \ref{noise_images}(d). In Figure (a) and (b), the FRC and the DPP fail for the case of the image with Gaussian noise; the SLaT (Figure (c)) and RSSFC (Figure (e)) misclassify the color of the area between the two neighbor circles; our proposed (Figure (f)) is better than the LC (Figure (d)) as it not only tells the five different colors but also preserves the shape of the circle.}
	\label{cleansixphases}
\end{figure*}


\begin{table*}[htb]
	\caption{The comparisons of SA and CPU-time in seconds by different methods in Figure \ref{testimages2}. \\(The best results in the simulation are given in bold.)}
	\centering
	\setlength{\tabcolsep}{1mm}{
		\scriptsize
		\begin{tabular}{c|llr|c|llr|c|llr}
			\hline
			\multicolumn{1}{l|}{Images} & Methods  & SA              & Time(s)       & \multicolumn{1}{l|}{Images} & Methods  & SA              & Time(s)       & \multicolumn{1}{l|}{Images} & Methods  & SA              & Time(s)       \\ \hline
			\multirow{8}{*}{Airplane}   & FRC      & 0.8674          & \textbf{3.33} & \multirow{8}{*}{Hill}       & FRC      & 0.7690          & \textbf{2.95} & \multirow{8}{*}{Flowers}    & FRC      & 0.7965          & \textbf{4.33} \\
			& DPP      & 0.6807          & 4.89          &                             & DPP      & 0.8646          & 4.80          &                             & DPP      & 0.5986          & 5.91          \\
			& SLaT     & 0.9067          & 7.82          &                             & SLaT     & 0.8891          & 9.77          &                             & SLaT     & 0.8997          & 5.58          \\
			& LC       & 0.9095          & 72.11         &                             & LC       & 0.9065          & 73.04         &                             & LC       & 0.9038          & 90.06         \\
			& CKC      & 0.9101          & 3546.64       &                             & CKC      & 0.9099          & 3094.92       &                             & CKC      & 0.9048          & 3915.16       \\
			& CQaS     & 0.9059          & 2109.49       &                             & CQaS     & 0.9056          & 2098.49       &                             & CQaS     & 0.9104          & 2589.49       \\
			& RSSFC    & 0.8754          & 44.61         &                             & RSSFC    & 0.7805          & 54.99         &                             & RSSFC    & 0.8505          & 71.93         \\
			& Proposed & \textbf{0.9170} & 14.34         &                             & Proposed & \textbf{0.9109} & 16.05         &                             & Proposed & \textbf{0.9151} & 16.55         \\ \hline
			\multirow{8}{*}{Parrot}     & FRC      & 0.8032          & \textbf{4.47} & \multirow{8}{*}{Ladybug}    & FRC      & 0.8245          & 2.00          & \multirow{8}{*}{Sunflowers} & FRC      & 0.8216          & 1.51          \\
			& DPP      & 0.5846          & 6.10          &                             & DPP      & 0.6940          & 1.55          &                             & DPP      & 0.6712          & 1.48          \\
			& SLaT     & 0.7539          & 6.08          &                             & SLaT     & 0.7984          & \textbf{1.51} &                             & SLaT     & 0.8568          & \textbf{1.30} \\
			& LC       & 0.8376          & 79.22         &                             & LC       & 0.8582          & 25.36         &                             & LC       & 0.8769          & 22.23         \\
			& CKC      & 0.7827          & 2040.88       &                             & CKC      & 0.8389          & 2576.99       &                             & CKC      & 0.8798          & 2191.62       \\
			& CQaS     & 0.8269          & 1135.65       &                             & CQaS     & 0.8513          & 1424.72       &                             & CQaS     & 0.8809          & 1264.82       \\
			& RSSFC    & 0.6904          & 86.41         &                             & RSSFC    & 0.7225          & 20.27         &                             & RSSFC    & 0.6387          & 16.86         \\
			& Proposed & \textbf{0.8389} & 15.52         &                             & Proposed & \textbf{0.8624} & 7.55          &                             & Proposed & \textbf{0.8824} & 7.05          \\ \hline
	\end{tabular}}\label{Table2}
\end{table*}
\section{The proposed model}

Now, we propose our model as follows
\begin{equation}\label{ourmodel}
	\min E(z)=\underset{z \in \mathcal{A}^{\Omega}}\min\left\{\lambda\|\nabla z\|_{1}+\frac{\mu}{2}\|\nabla z\|_{2}^{2}+
	\langle z, w\rangle\right\},
\end{equation}
where $\langle z, w\rangle$ is defined in Section \ref{The smoothed Potts model}.
 The set $\mathcal{A}$ is non-convex, so the problem (\ref{ourmodel}) is non-convex. Here we perform a convex relaxation, replacing $\mathcal{A}$ with its convex hull (the simplex $\Delta$), which is the set of vectors with nonnegative elements whose sum is 1. This strategy meets the goal (a): to avoid solving non-convex and combinatorial problem for the sake of stability, then we reformulate the problem (\ref{ourmodel}) as follows
 
\begin{equation}\label{theconstrainedproblem}
			\min _{z}\left\{\lambda\|\nabla z\|_{1}+\frac{\mu}{2}\|\nabla z\|_{2}^{2}+ \langle z, w\rangle+\imath_{\Delta}(z)\right\}.
\end{equation}

 Since model (\ref{theconstrainedproblem}) is convex, there are many methods to solve it such as the split-Bregman algorithm \cite{goldstein2009split}, which is used widely to solve many problems based on TV regularization and the alternating direction method of multipliers (ADMM) \cite{boyd2011distributed,figueiredo2010restoration,guo2016convergence}.
Specially, the Chambolle-Pock algorithm \cite{chambolle2011first} proves a convergence rate.

In the following, we use the Chambolle-Pock algorithm to solve (\ref{theconstrainedproblem}), first we introduce new variable $v=\nabla z$, then we consider the following primal-dual optimization problem

\begin{equation}\label{dualform}
	\begin{split}
		\min _{z,v} \max _{q}\bigg\{&\lambda\|v\|_{1}+\frac{\mu}{2}\|v\|_{2}^{2}+\langle z, w\rangle+\\&\langle v-\nabla z, q\rangle+\imath_{\Delta}(z)\bigg\}.
	\end{split}
\end{equation}
Then the Chambolle-Pock algorithm is defined through the following iterations
\begin{numcases}{}
	q^{(k+1)}=\arg \max _{q}\bigg\{\left\langle\overline{v}^{(k)}-\nabla \overline{z}^{(k)}, q\right\rangle\nonumber\\
	\qquad \qquad\qquad\qquad-\frac{1}{2 \sigma}\left\|q-q^{(k)}\right\|_{2}^{2}\bigg\},\label{qsubproblem}\\                                 
	z^{(k+1)}=\arg \min _{z}\bigg\{-\left\langle\nabla z, q^{(k+1)}\right\rangle+\langle z,w \rangle+\imath_{\Delta}(z)\nonumber\\
	\qquad \qquad\qquad\qquad+\frac{1}{2 \tau}\left\|z-z^{(k)}\right\|_{2}^{2}\bigg\},\label{zsubproblem}\\
	v^{(k+1)}=\arg \min _{v}\bigg\{\lambda\|v\|_{1}+\frac{\mu}{2}\|v\|_{2}^{2}+\left\langle v, q^{(k+1)}\right\rangle\nonumber\\
	\qquad \qquad\qquad\qquad+\frac{1}{2 \tau}\left\|v-v^{(k)}\right\|_{2}^{2}\bigg\},\label{vsubproblem}\\
	\overline{z}^{(k+1)}=2 z^{(k+1)}-z^{(k)},\label{zbar}\\
	\overline{v}^{(k+1)}=2 v^{(k+1)}-v^{(k)}\label{vbar}.
\end{numcases}

Since the objective functions (\ref{qsubproblem}) is quadratic,
the update of $q$ can be computed efficiently
\begin{equation}\label{qsubproblemsolution}
	q^{(k+1)}=\sigma\left(\overline{v}^{(k)}-\nabla \overline{z}^{(k)}\right)+q^{(k)}.
\end{equation}
The solution of (\ref{vsubproblem}) can be easily obtained by applying
the soft thresholding operator. Denoting $t^{(k)}=\frac{\tau}{\mu \tau+1}\left(\frac{1}{\tau} v^{(k)}-q^{(k+1)}\right)$, we have
\begin{equation}\label{v_xsolution}
	v_{x}^{(k+1)}=\max \left\{\left|t^{(k)}\right|-\frac{\lambda\tau}{\mu \tau+1}, 0\right\} \cdot \frac{t_{x}^{(k)}}{\left|t^{(k)}\right|},
\end{equation}
\begin{equation}\label{v_ysolution}
	v_{y}^{(k+1)}=\max \left\{\left|t^{(k)}\right|-\frac{\lambda\tau}{\mu \tau+1}, 0\right\} \cdot \frac{t_{y}^{(k)}}{\left|t^{(k)}\right|}.
\end{equation}
Denoting $z_0^{k}=z^{k}-\tau(w+\operatorname{div} q^{(k+1)})$, we have
\begin{equation}
	z^{(k+1)}=\arg \min _{z}\left\{\frac{1}{2 \tau}\left\|z-z_0^{k}\right\|_{2}^{2}+\imath_{\Delta}(z)\right\},
\end{equation} then
\begin{equation}\label{zsolution}
	z^{(k+1)}=P_{\Delta}(z_0^{k}),
\end{equation}
where $P_{\Delta}$ is the simplex projection, which could be computed efficiently by the code from \url{https://lcondat.github.io/}.

Our one-stage method has the following three steps:
\begin{itemize}
	\item [1.]  First, we use the hill-climbing procedure to detect the number of segments ($K$) and use the K-means algorithm to get $K$ colors $c_{k}$.
	\item [2.]  Second, we solve our one-stage model (merging the process of smoothing and segmentation) via the Chambolle-Pock algorithm to get the assignment array $z$.
	\item [3.]  Third, we get the segmented image $u$ via simple summation.
\end{itemize}

Then the whole process of our one-stage strategy image segmentation method for clean or noisy images can be summarized by the following Algorithm 1.
\begin{table}[htbp]
	\centering
	\begin{tabular}{l}
		\toprule
		\textbf{Algorithm 1:} Chambolle-Pock algorithm to solve (\ref{theconstrainedproblem})\\
		\midrule
		1. Initialize: given the $z^{0},\overline {z}^{0} $, $c_{k}$, $q^{0}=0$, $v^{0}=\overline {v}^{0}=0$.\\
		2. Do k=0, 1,..., maxiteation:\\
		\quad a. compute $q^{k+1}$ by(\ref{qsubproblemsolution});\\
		\quad b. compute $v^{k+1}$ by (\ref{v_xsolution}) and (\ref{v_ysolution});\\
		\quad c. compute $z^{k+1}$ by  (\ref{zsolution});\\
		\quad d. update $\overline{z}^{k+1}$ and $\overline{v}^{k+1}$ by (\ref{zbar}) and (\ref{vbar}).\\
		3. Output: the segmented $u$ by (\ref{segmentedu}).\\
		\bottomrule
	\end{tabular}
\end{table}

\begin{figure*}[htb]
	\centering
	\subfigure[FRC \cite{li2010multiphase}]{
		\includegraphics[width=1.in]{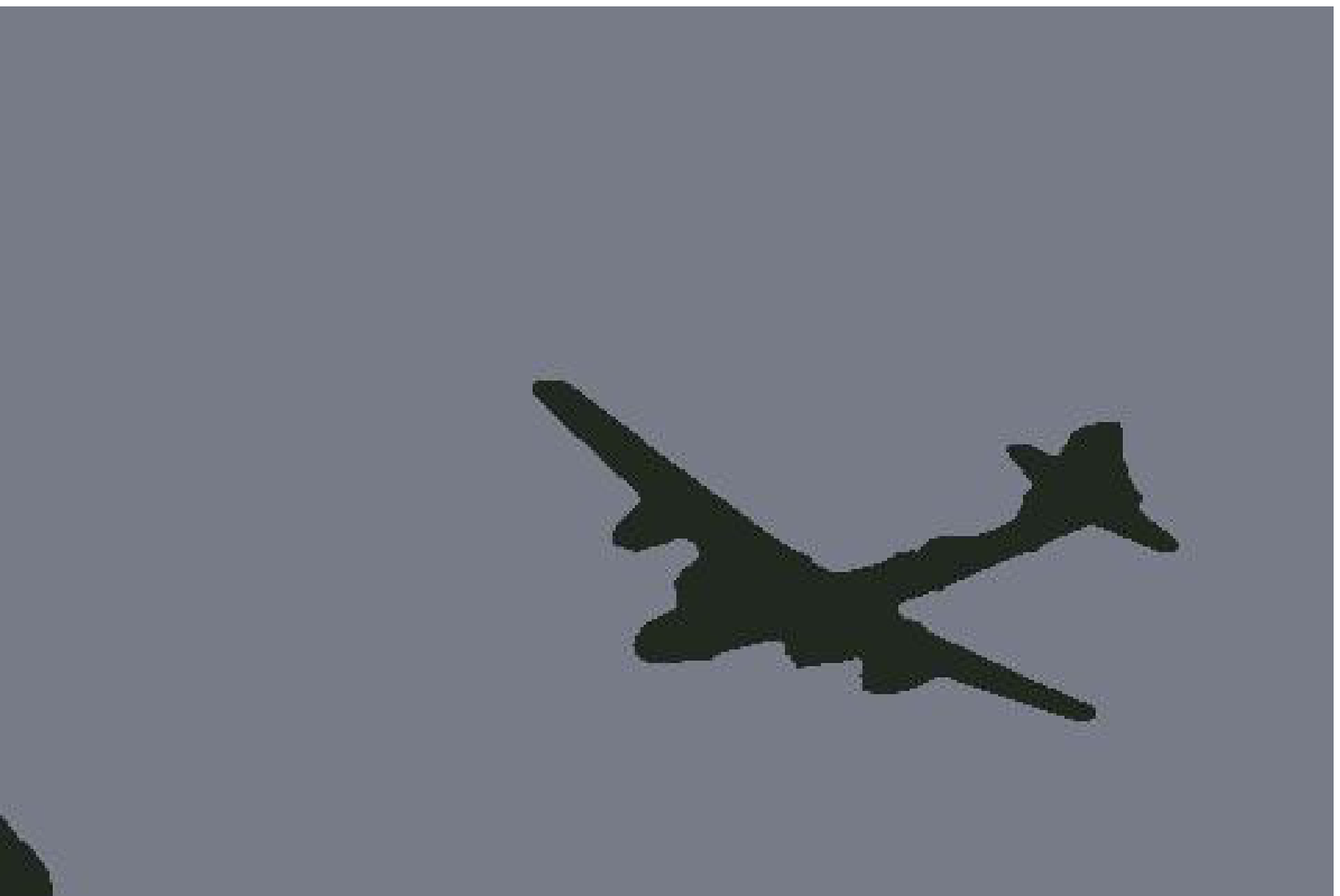}}
	\subfigure[DPP \cite{storath2014fast}]{
		\includegraphics[width=1.in]{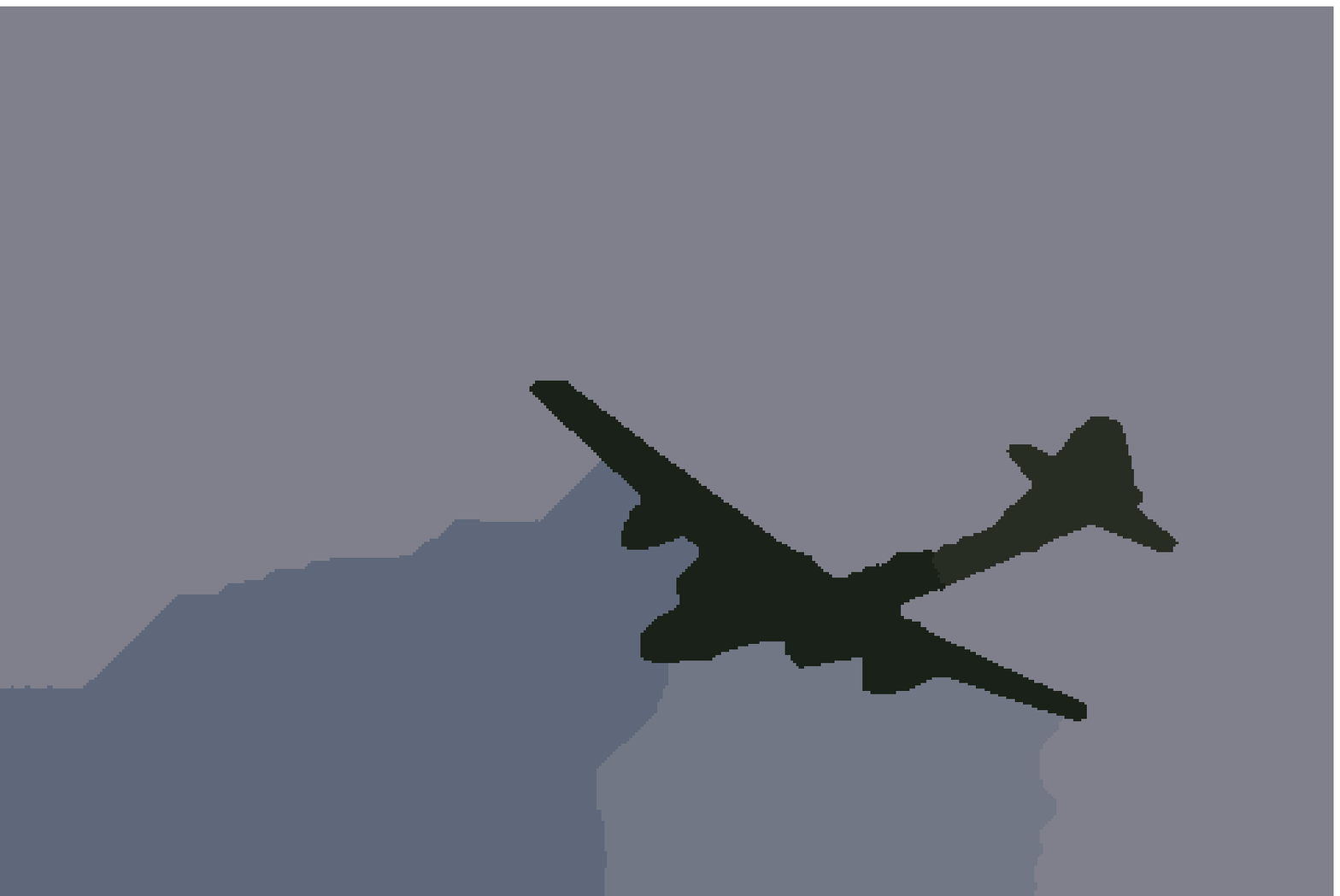}}
	\subfigure[SLaT \cite{cai2017three}]{
		\includegraphics[width=1.in]{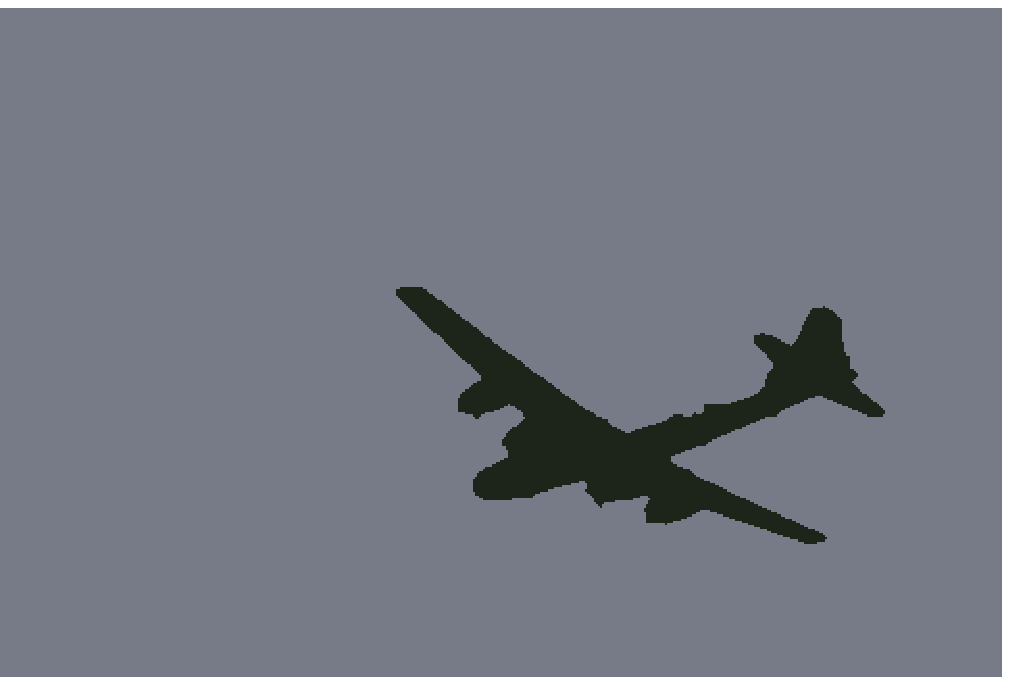}}
	\subfigure[LC \cite{condat2017discrete}]{
		\includegraphics[width=1.in]{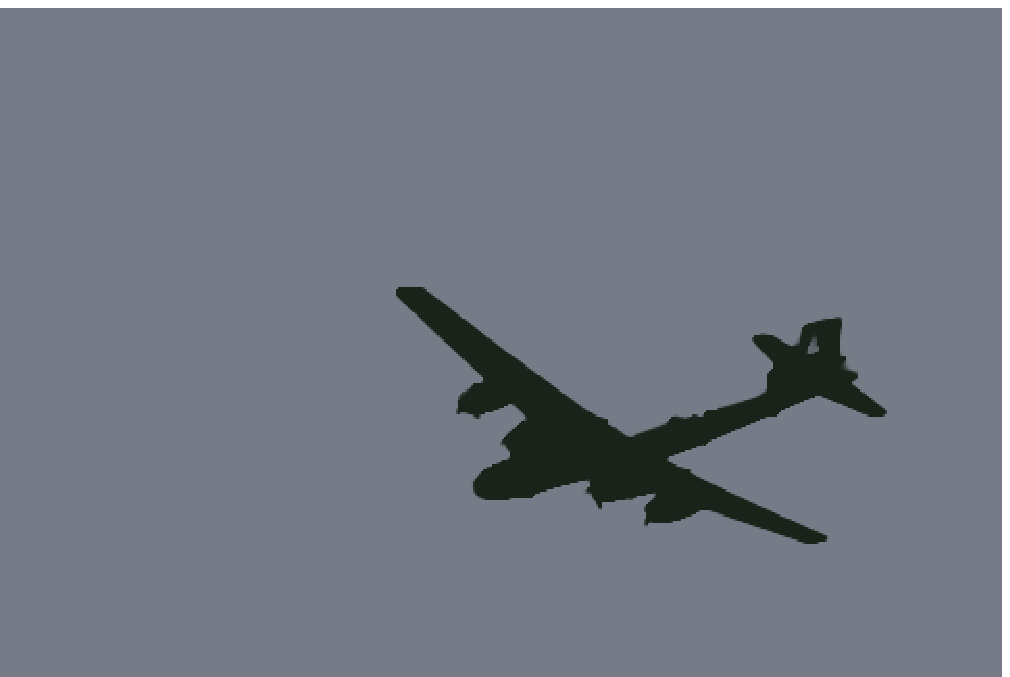}}
	\subfigure[RSSFC \cite{jia2020robust}]{	
		\includegraphics[width=1.in]{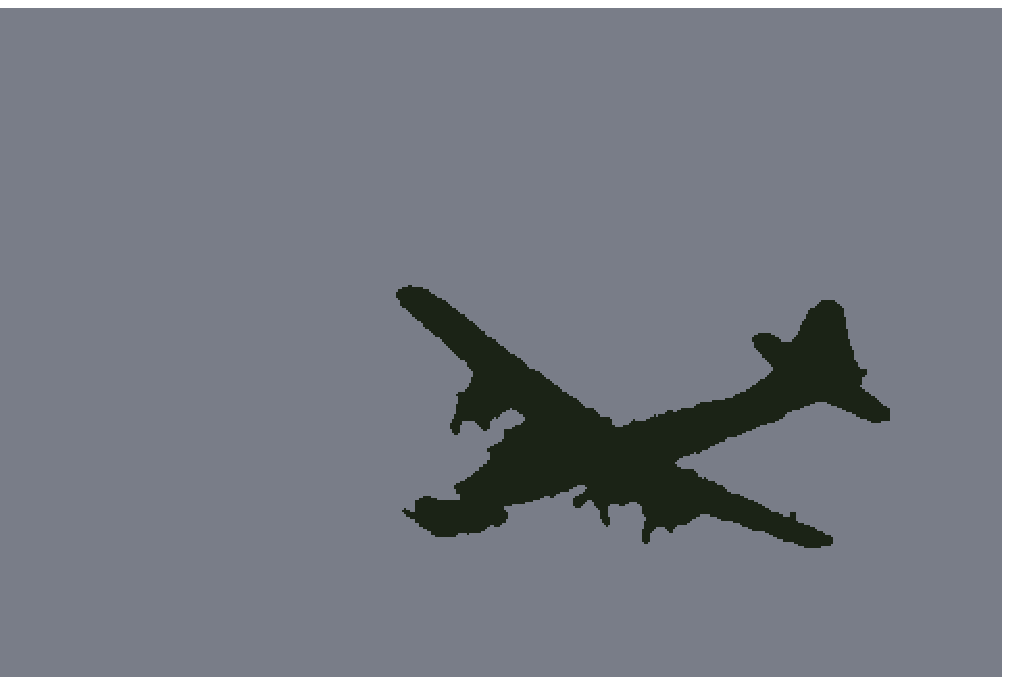}}
	\subfigure[Proposed]{
		\includegraphics[width=1in]{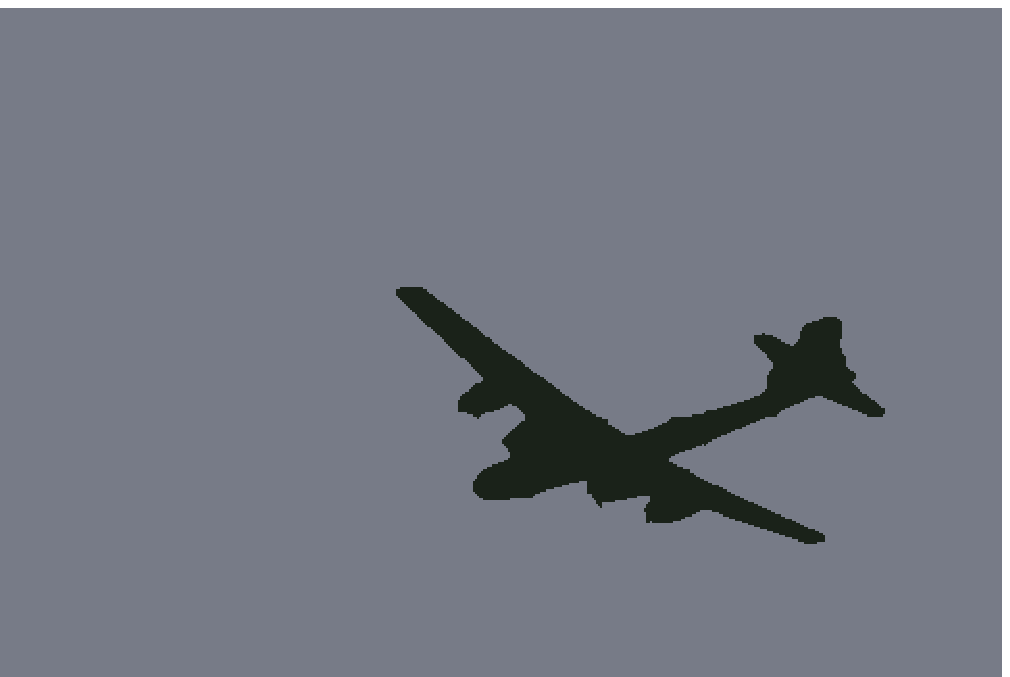}}
	\caption{The segmentation results of Airplane in Figure \ref{testimages2}(a). The FRC (Figure (a)) and the RSSFC (Figure (e)) do not clearly cut the airplane out as there is a small area that is not separated into the background; the DPP (Figure (b)) has a poor performance on this task; The SLaT (Figure (c)), the LC (Figure (d)) and our proposed (Figure (f), $\lambda=0.2$) cut the plane from the sky (background), and segment the image into two reasonable parts.}
	\label{cleanairplane}
\end{figure*}

\begin{figure*}[htb]
	\centering	
	
	\subfigure[FRC \cite{li2010multiphase}]{
		\includegraphics[width=1.in]{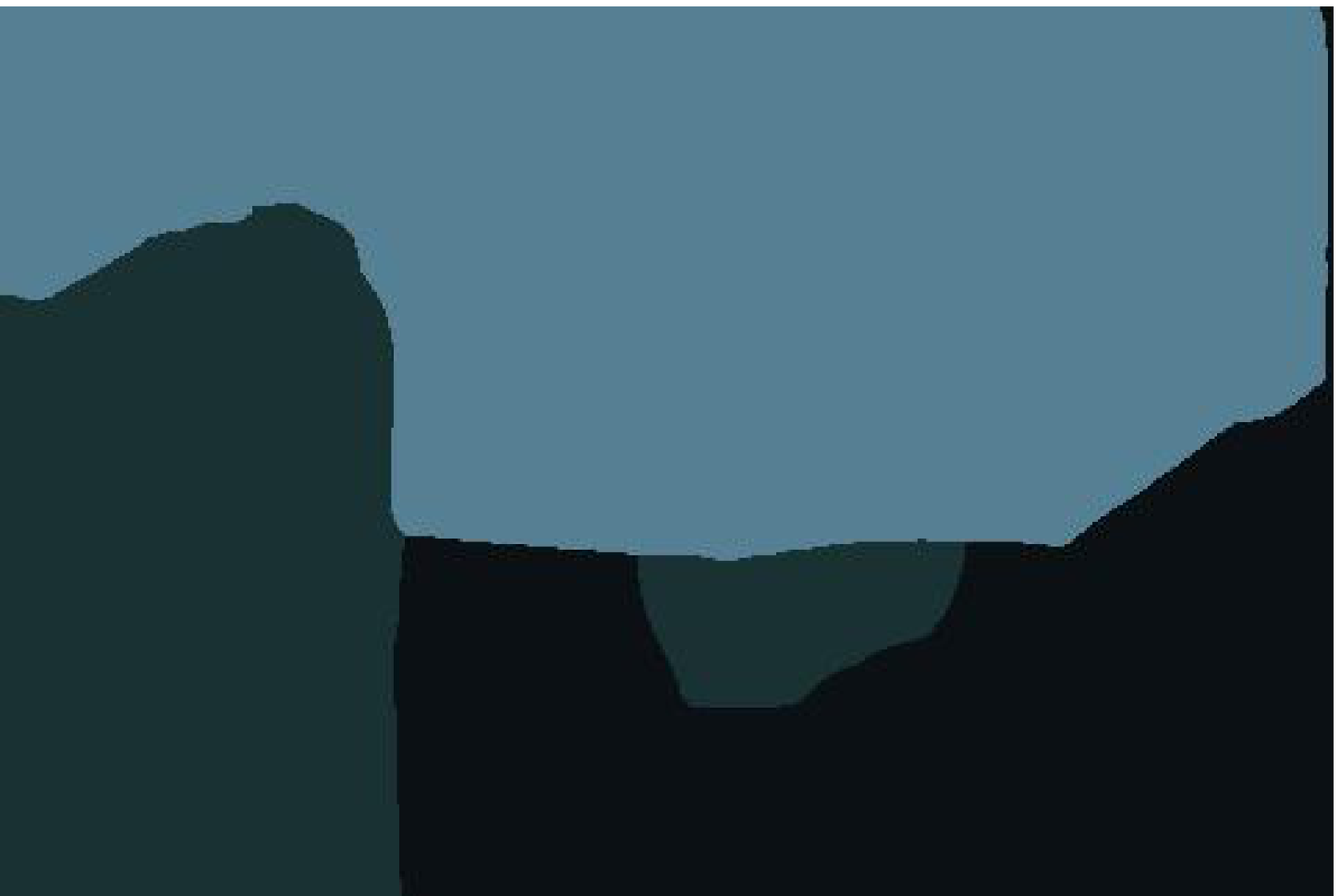}}
	\subfigure[DPP \cite{storath2014fast}]{
		\includegraphics[width=1.in]{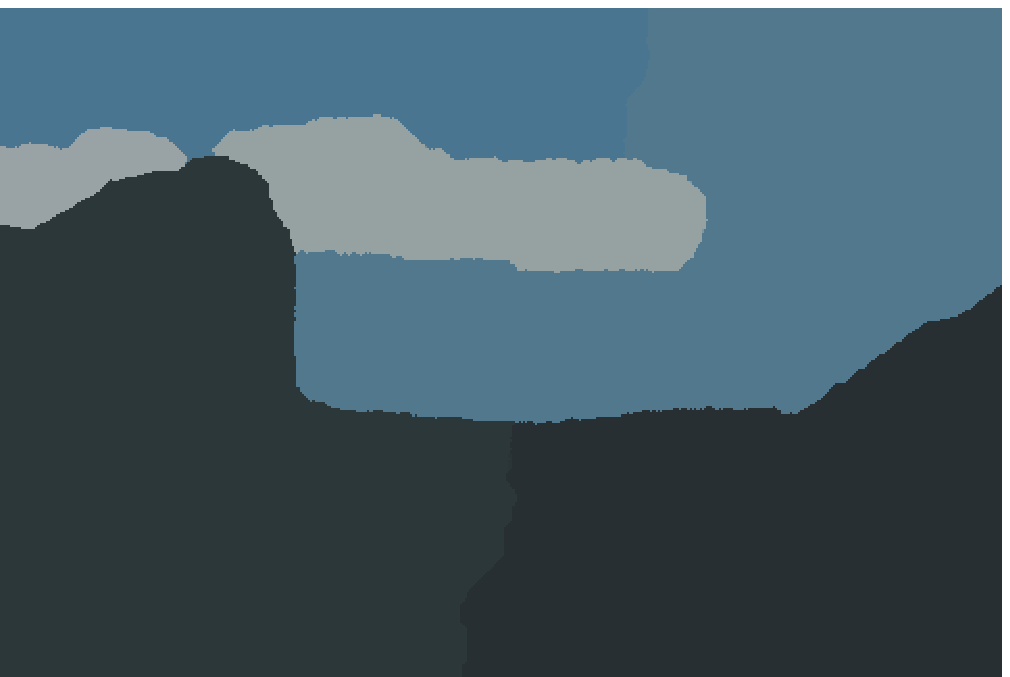}}
	\subfigure[SLaT \cite{cai2017three}]{
		\includegraphics[width=1.in]{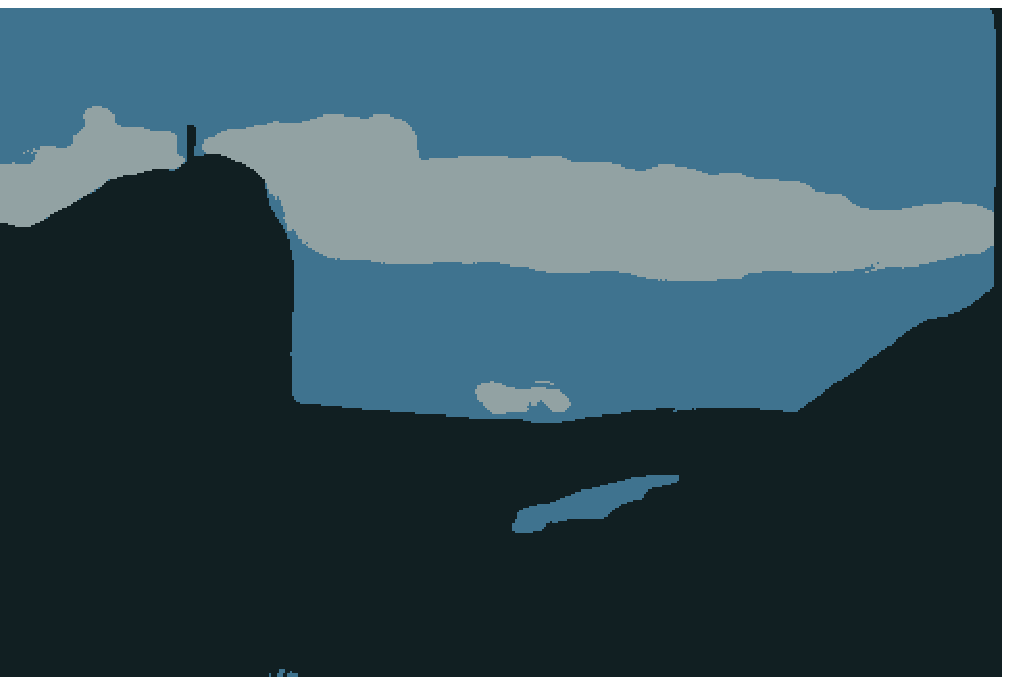}}
	\subfigure[LC \cite{condat2017discrete}]{
		\includegraphics[width=1.in]{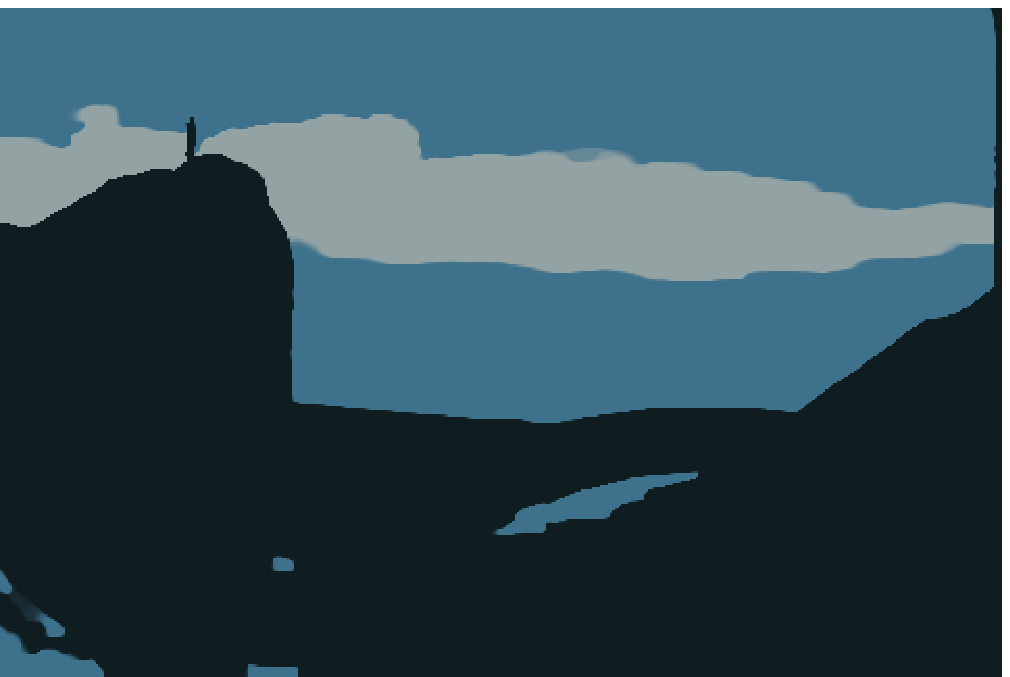}}
	\subfigure[RSSFC \cite{jia2020robust}]{	
		\includegraphics[width=1.in]{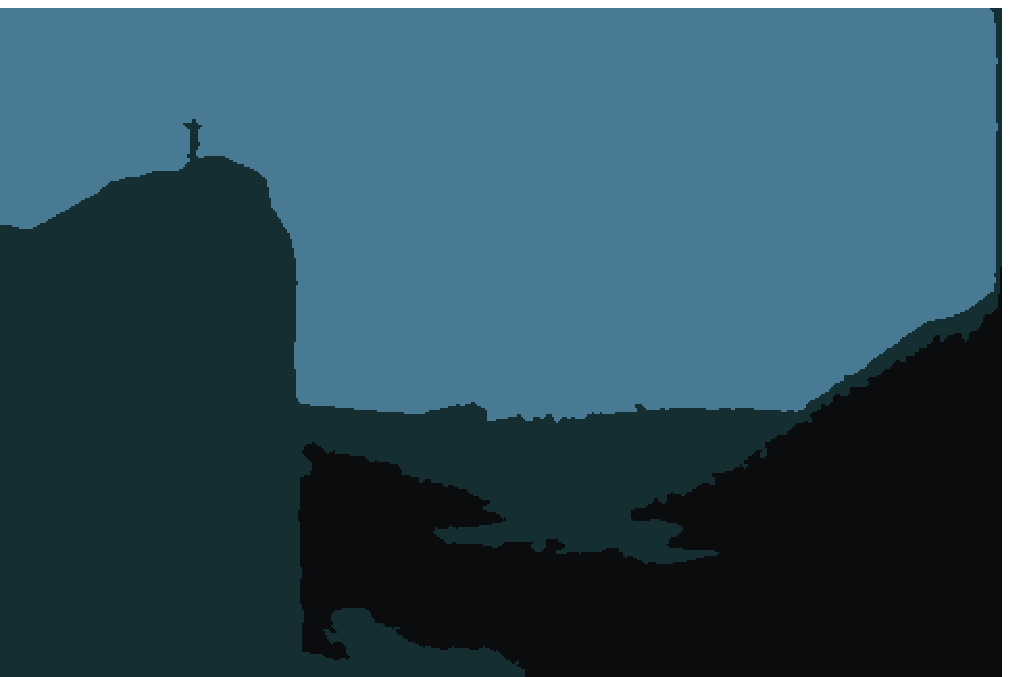}}
	\subfigure[Proposed]{
		\includegraphics[width=1in]{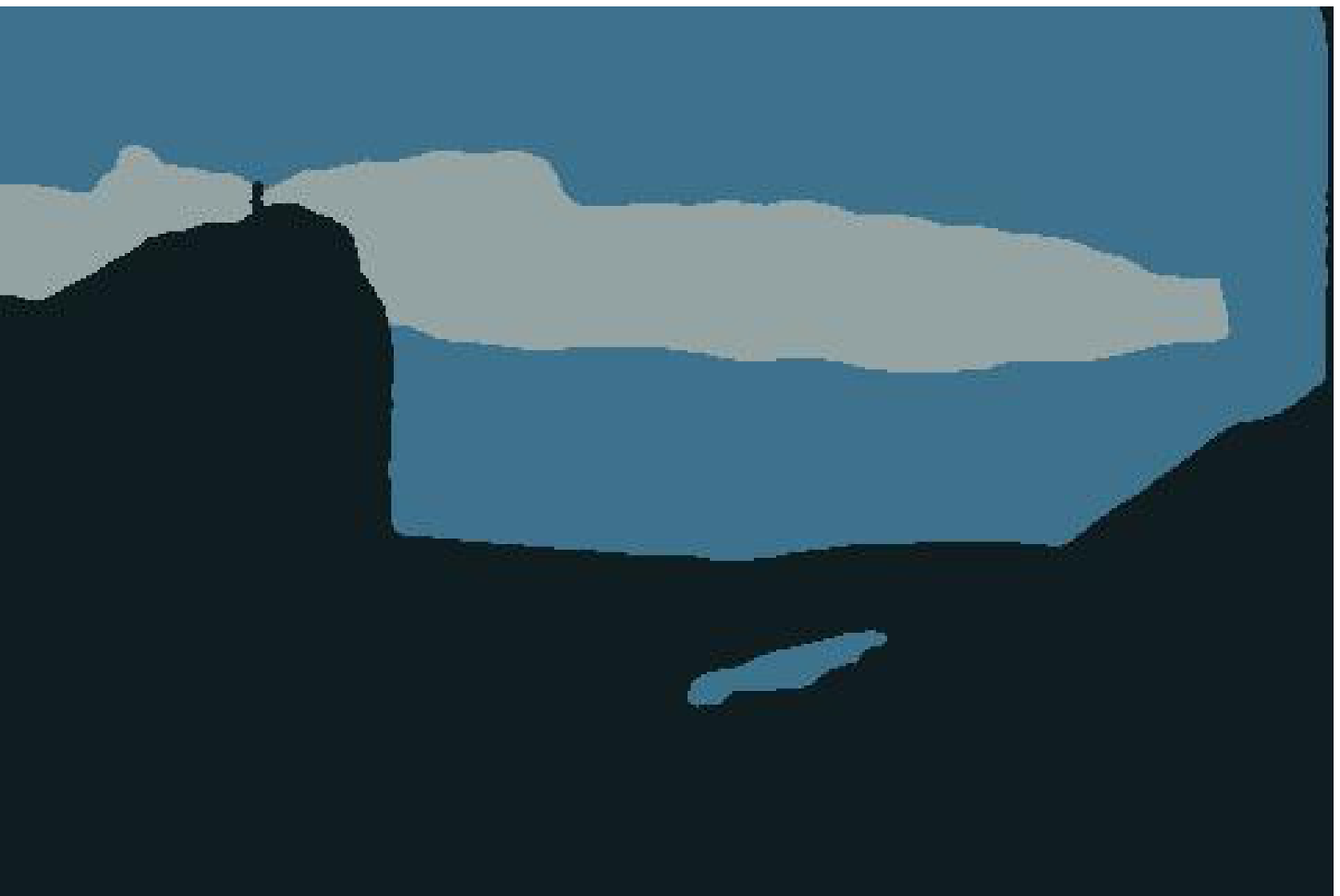}}
	\caption{The segmentation results of Hill in Figure \ref{testimages2}(b). The FRC (Figure (a)) and RSSFC (Figure(e)) fail to separate the hills as an entirety; the DPP (Figure(b)) is somewhat under-segmented that some small features have vanished; the LC (Figure (d)) does not cut the hill clearly out; the SLaT (Figure (c)) and our proposed (Figure (f), $\lambda=0.2$) get relatively better results.}
	\label{cleanhill}
\end{figure*}

\begin{figure*}[htb]
	\centering
	\subfigure[FRC \cite{li2010multiphase}]{
		\includegraphics[width=1in]{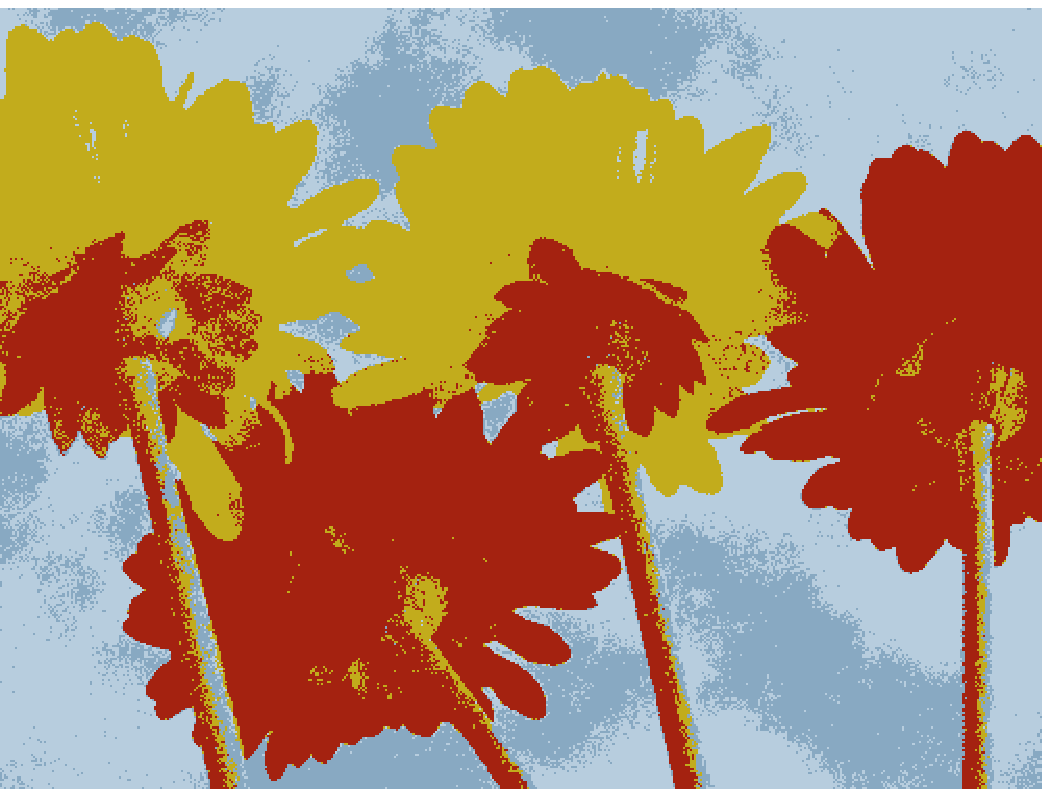}}
	\subfigure[DPP \cite{storath2014fast}]{
		\includegraphics[width=1.in]{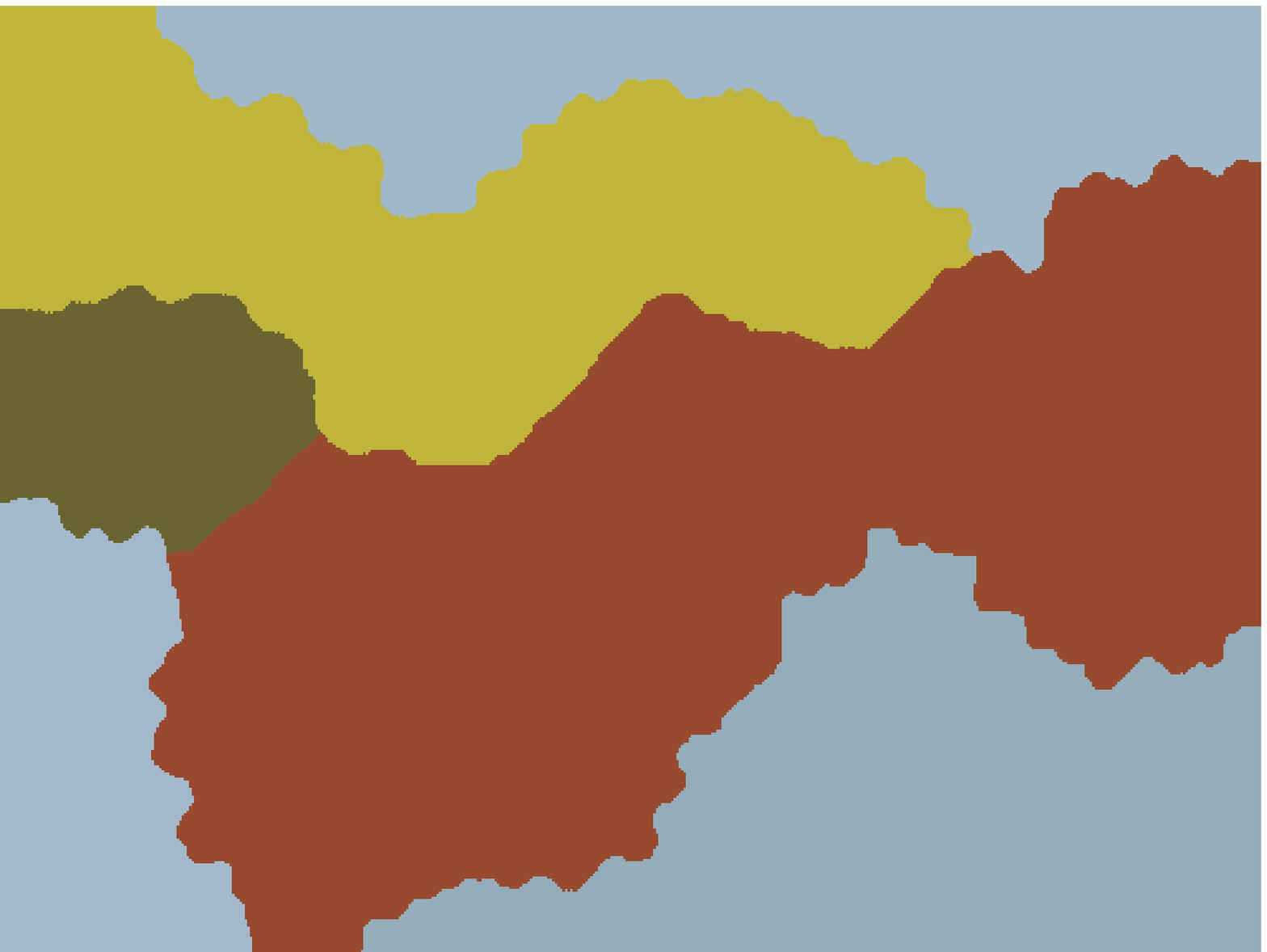}}
	\subfigure[SLaT \cite{cai2017three}]{
		\includegraphics[width=1.in]{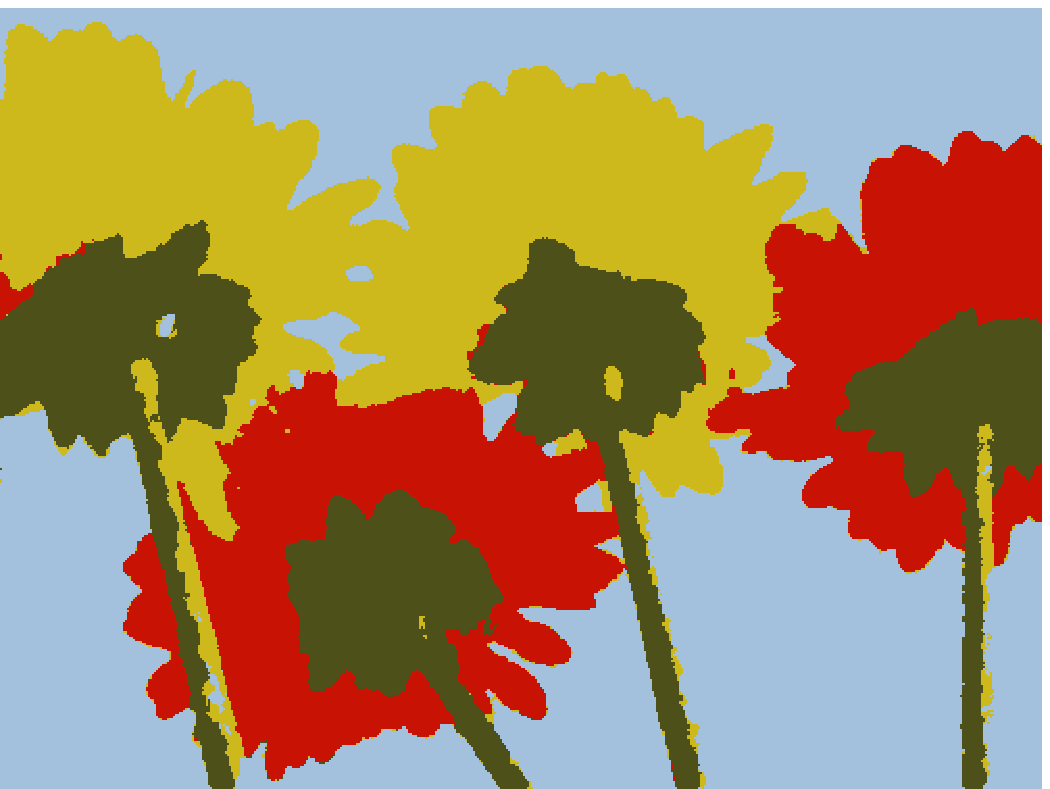}}
	\subfigure[LC \cite{condat2017discrete}]{
		\includegraphics[width=1.in]{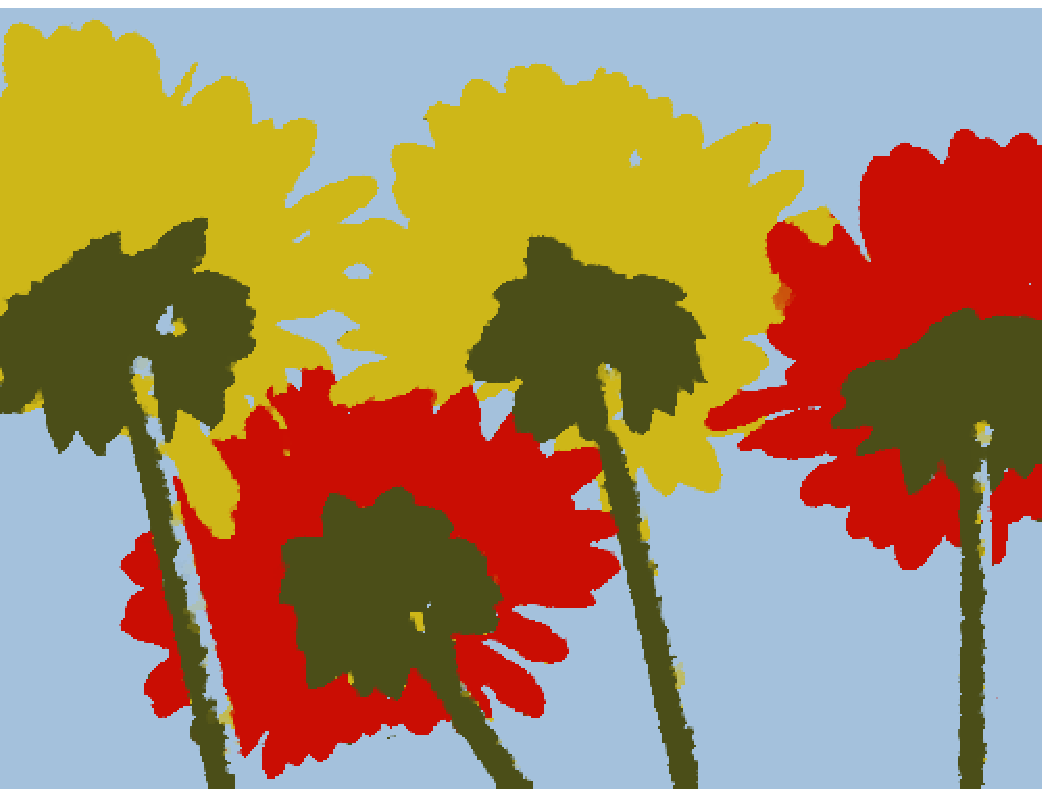}}
	\subfigure[RSSFC \cite{jia2020robust}]{	
		\includegraphics[width=1.in]{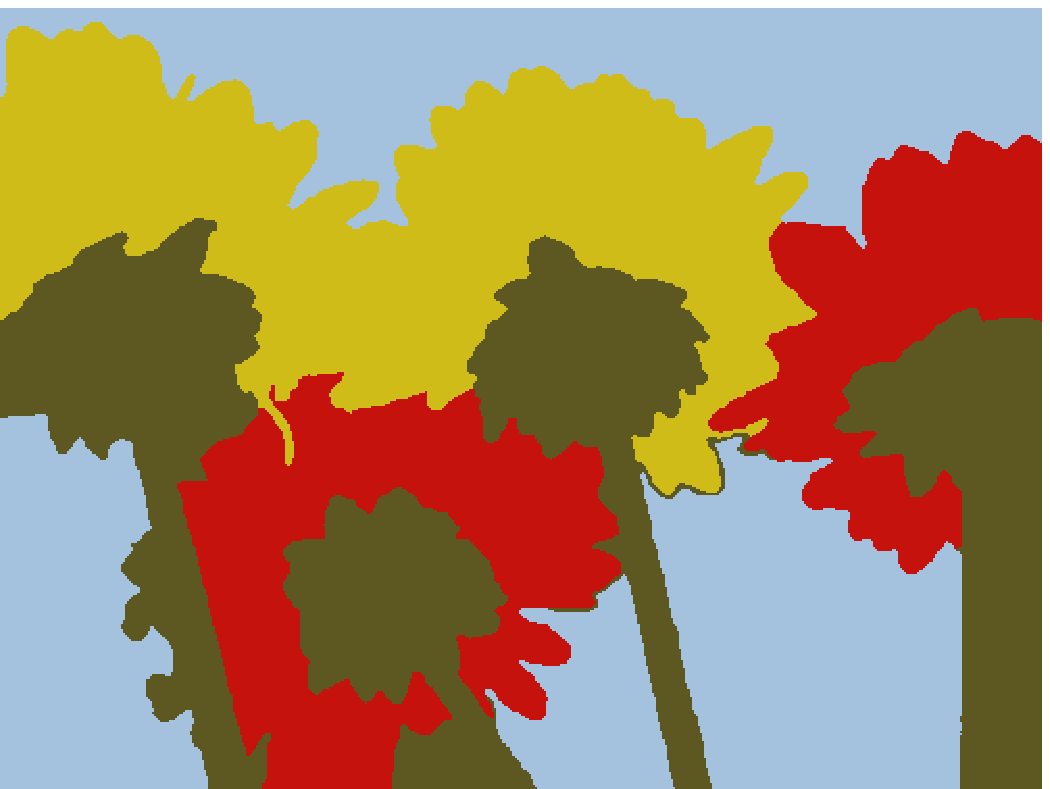}}
	\subfigure[Proposed]{
		\includegraphics[width=1.in]{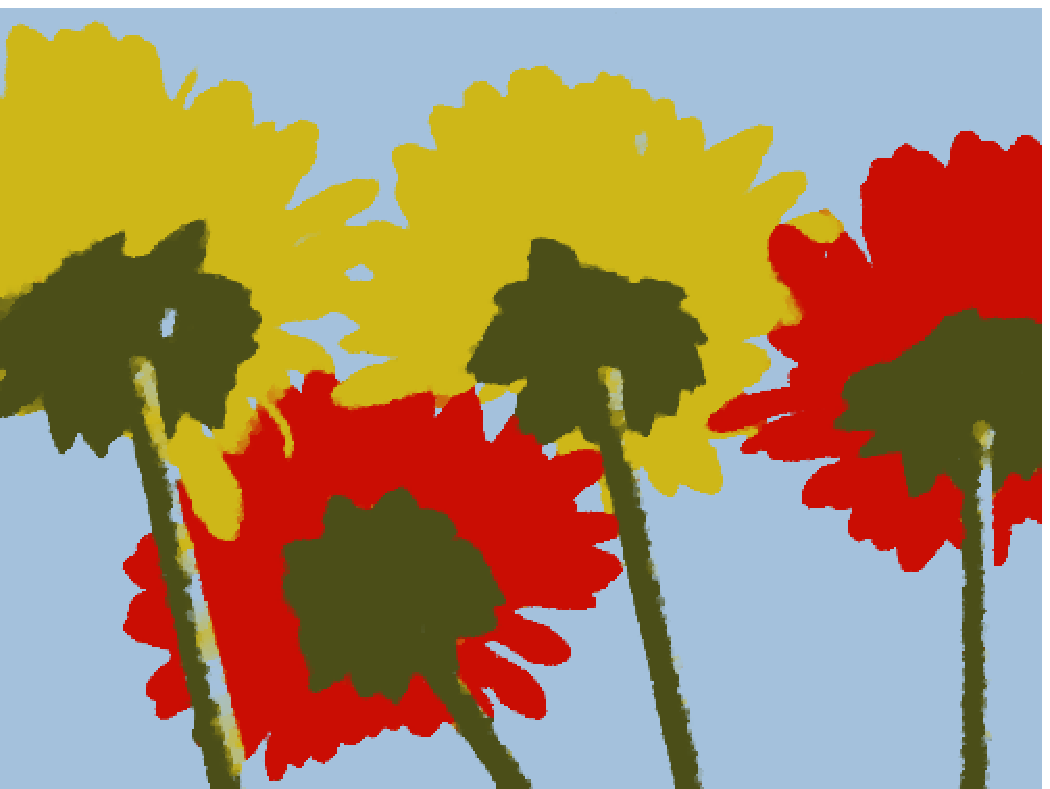}}
	\caption{The segmentation results of Flowers in Figure \ref{testimages2}(c) with Gaussian noise ($\tilde{\mu}=0$, $\sigma^2=0.1$). The FRC (Figure (a)) fails to distinguish the colors; the DPP (Figure (b)) does not make out the objects in the flowers; the SLaT (Figure (c)) misclassifies the area around the receptacle of the first and third flowers; the LC (Figure (d)) produces some yellow ``dots''; the RSSFC (Figure (e)) misclassifies the second flowers; our proposed (Figure (f), $\lambda=0.4$) gets a relatively better result.}
	\label{cleanflowers}
\end{figure*}

\begin{figure*}[htb]
	\centering
	\includegraphics[width=4.0in]{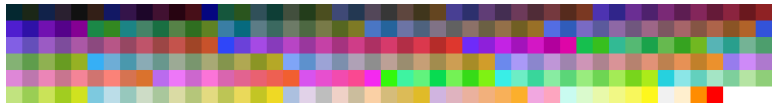}
	\caption{The 279 colors candidates in \cite{condat2017convex} obtained by sampling the CIELAB space.
	}
	\label{CKClabels}
\end{figure*}

\begin{figure*}[htb]
	\centering
	
	\subfigure[SLaT \cite{cai2017three}]{
		\includegraphics[width=1.in]{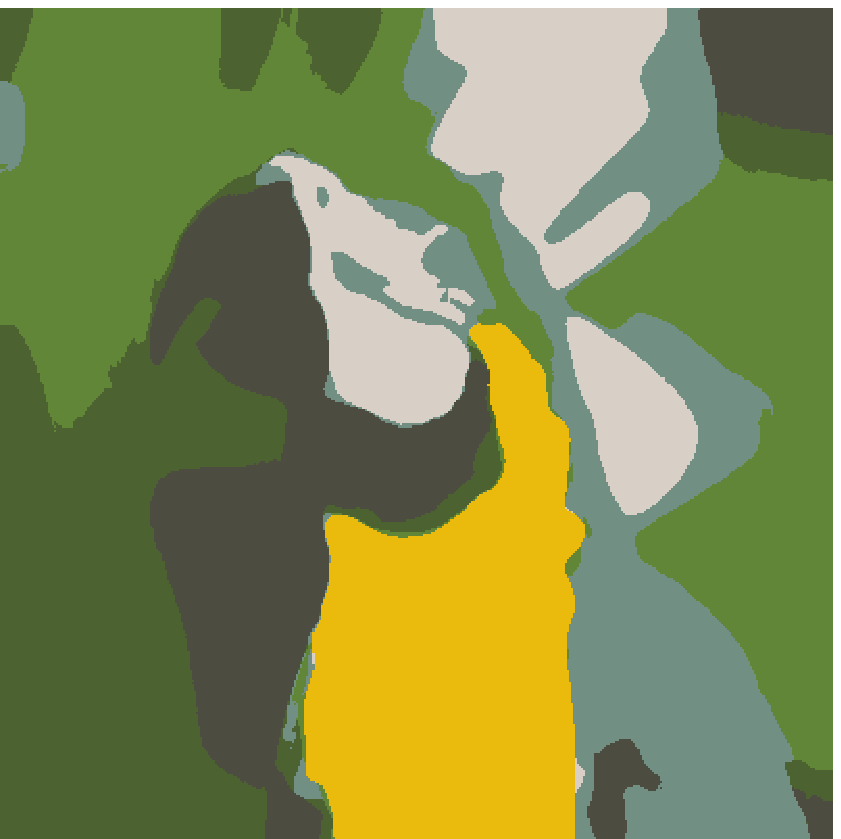}}
	\subfigure[LC \cite{condat2017discrete}]{
		\includegraphics[width=1.in]{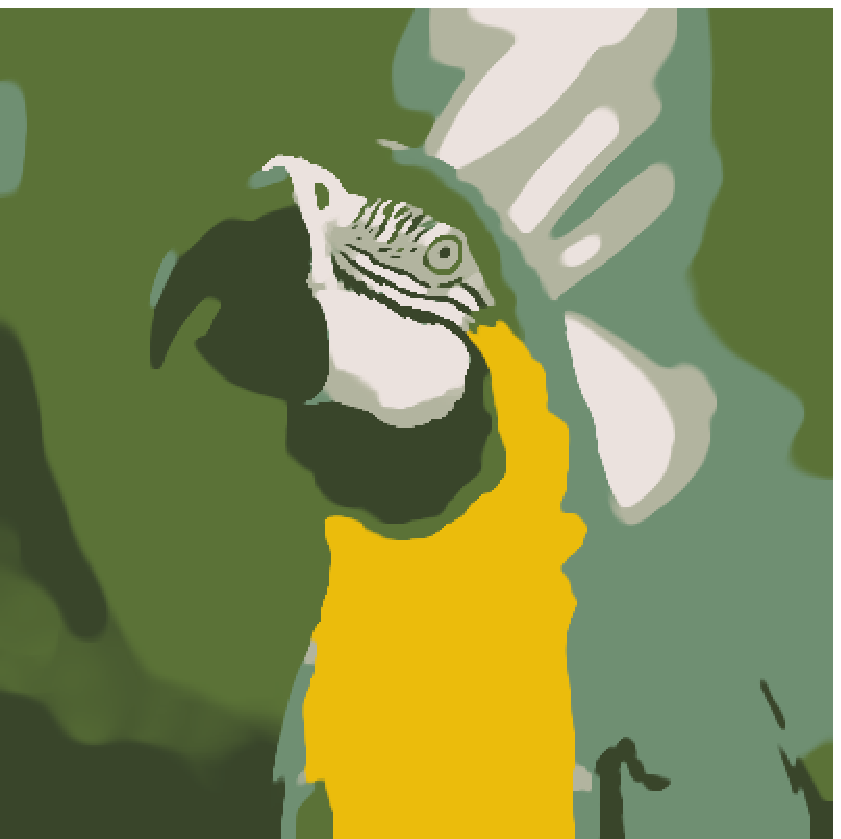}}
	\subfigure[CKC \cite{condat2017convex}]{
		\includegraphics[width=1.in]{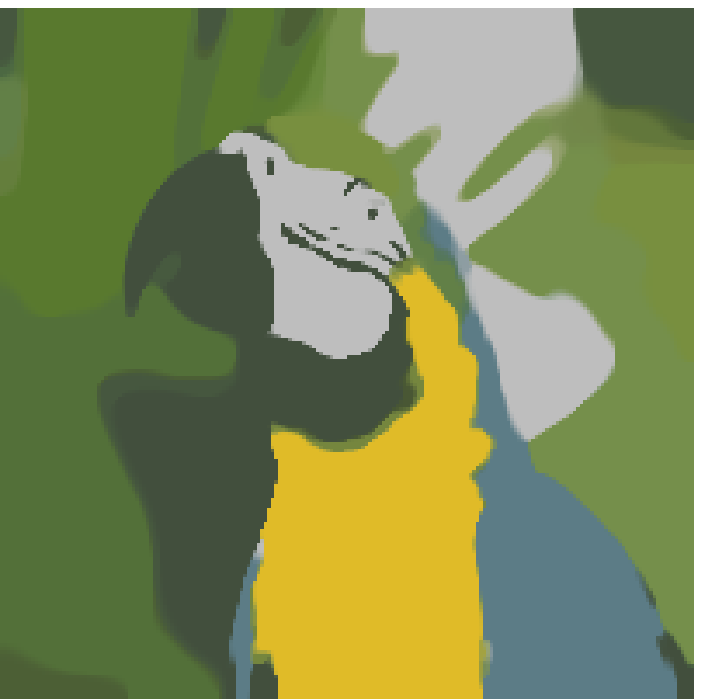}}
	\subfigure[CQaS \cite{condat2017convex}]{
		\includegraphics[width=1.in]{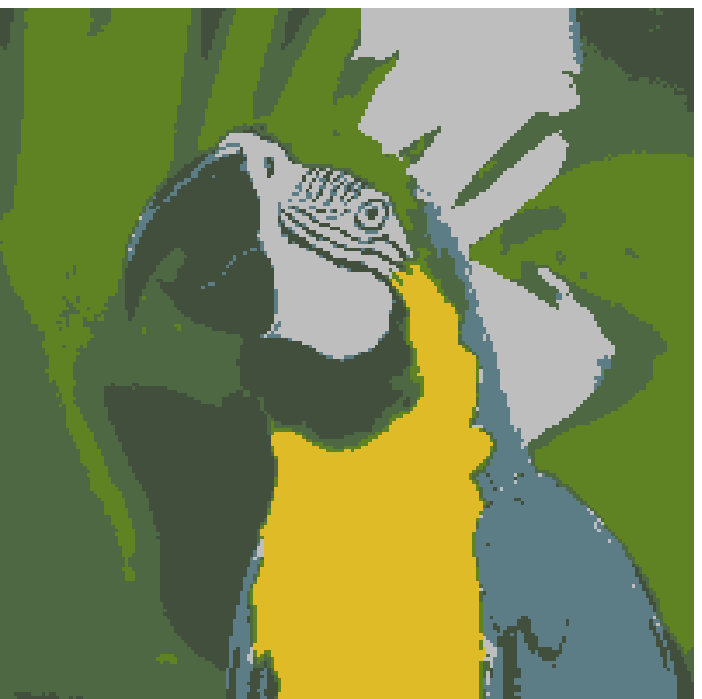}}
	\subfigure[RSSFC \cite{jia2020robust}]{	
		\includegraphics[width=1.in]{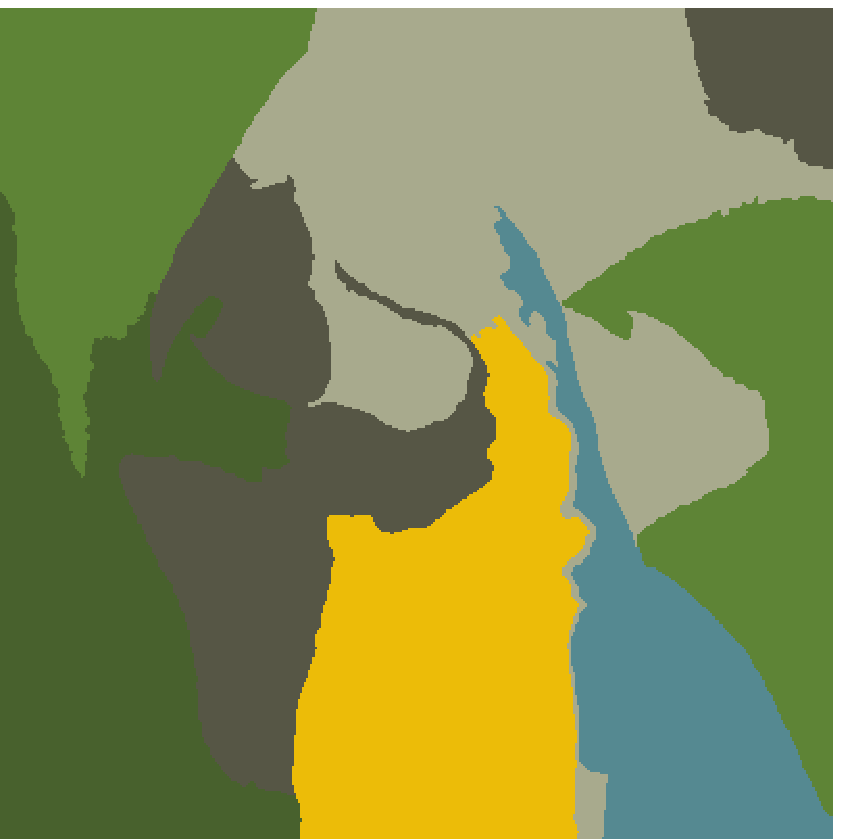}}
	\subfigure[ Proposed]{
		\includegraphics[width=1.in]{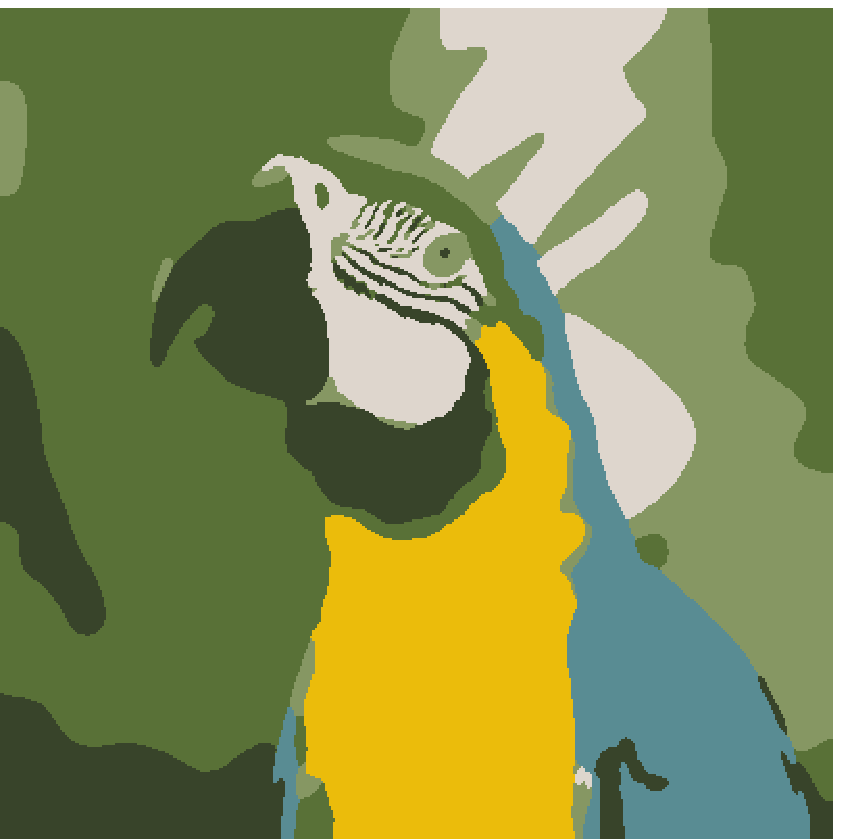}}
	\caption{The segmentation results of Parrot in Figure \ref{testimages2}(d). We test the K-means palette to specify the color set shown in Figure \ref{cleansixphases_p}(d). The SLaT in Figure (a) and CKC in Figure (c) undercut the image; the LC in Figure (b) gets a more smooth result than the former methods; the CQaS in Figure (d) and RSSFC in Figure (e) have a poor performance obviously; our proposed gets a relatively good result at the object (parrot) in Figure (f) ($\lambda=0.09$).}
	\label{cleanparrot}
\end{figure*}

\begin{figure*}[htb]
	\centering
	
	\subfigure[SLaT \cite{cai2017three}]{
		\includegraphics[width=1in]{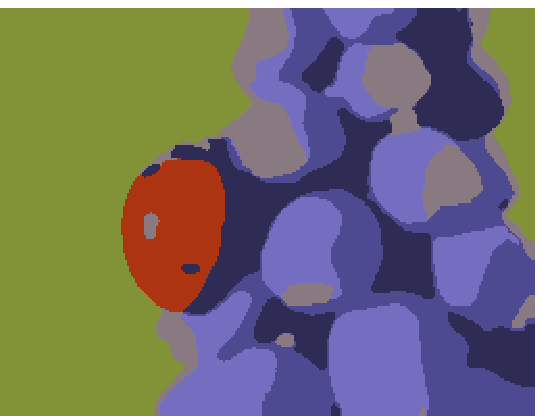}}
	\subfigure[LC \cite{condat2017discrete}]{
		\includegraphics[width=1.in]{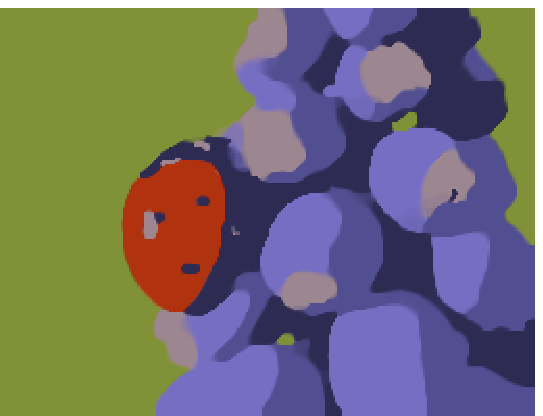}}
	\subfigure[CKC \cite{condat2017convex}]{
		\includegraphics[width=1.in]{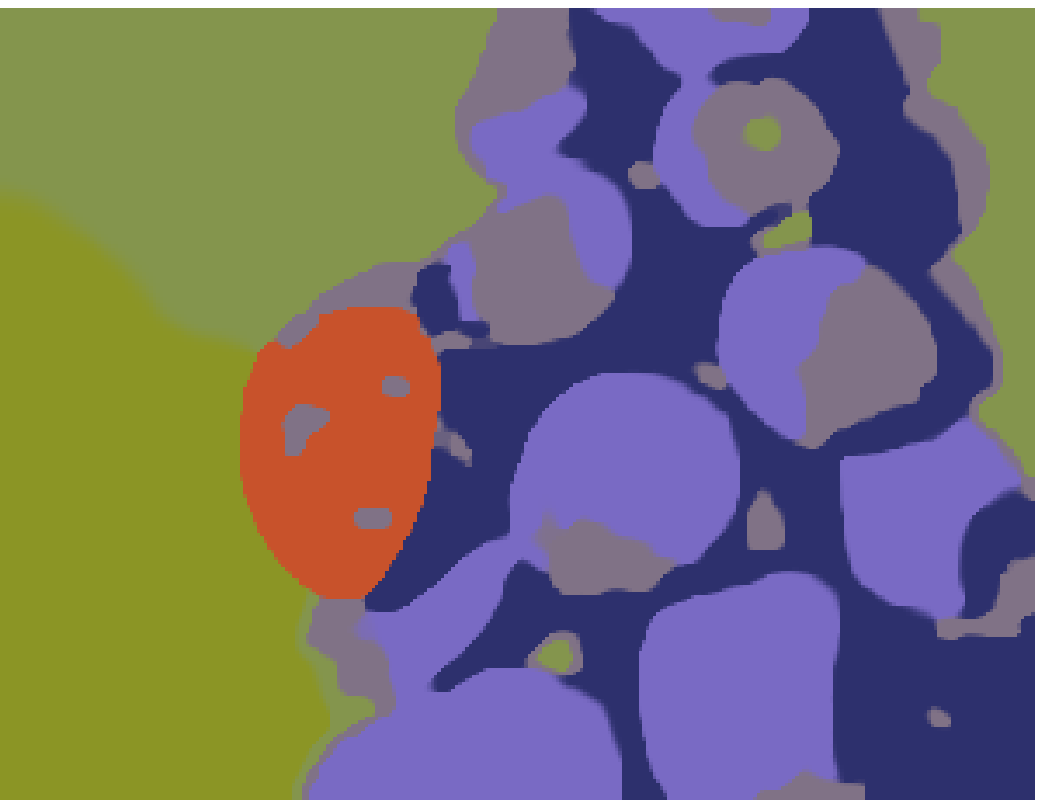}}
	\subfigure[CQaS \cite{condat2017convex}]{
		\includegraphics[width=1.in]{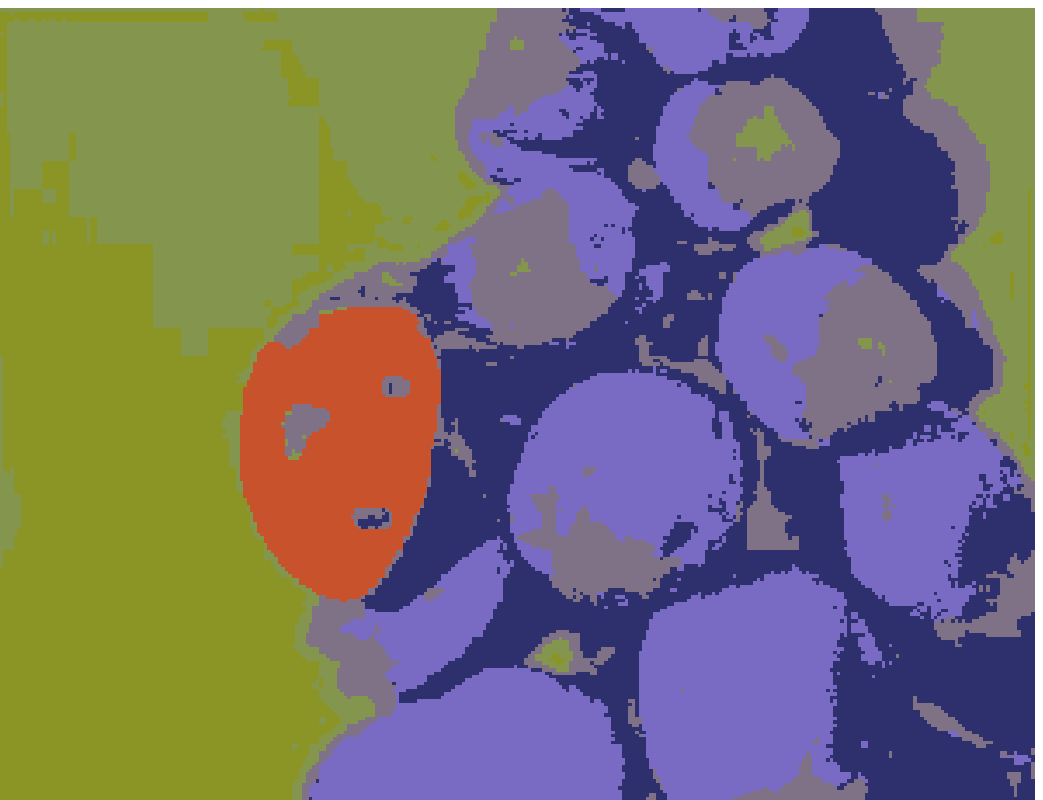}}
	\subfigure[RSSFC \cite{jia2020robust}]{	
		\includegraphics[width=1.in]{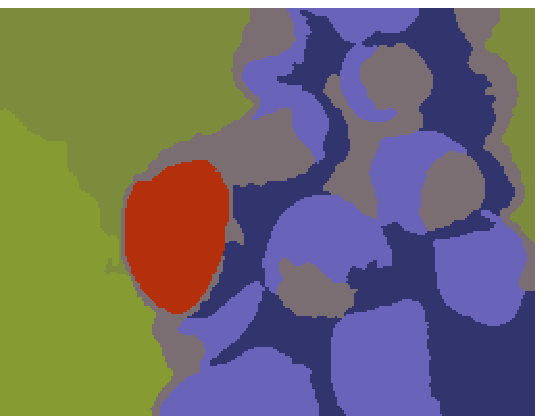}}
	\subfigure[Proposed]{
		\includegraphics[width=1in]{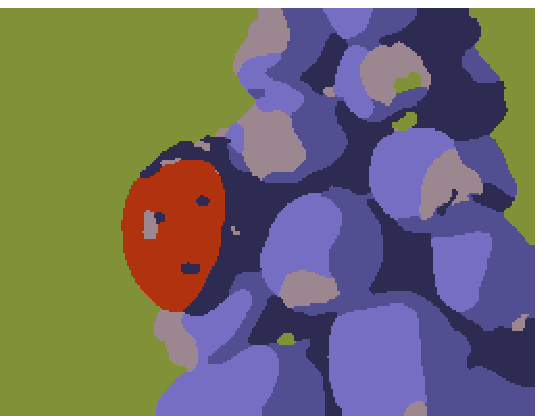}}
	\caption{Segmentation results of Ladybug in Figure \ref{testimages2}(e).  The SLaT (Figure (a)), LC (Figure (b)), CKC (Figure (c)) and our method (Figure (f), $\lambda=0.05$) all have a good result; the RSSFC (Figure (e)) has some under-segmentation to the objects in the image; our method is less complicated than the CKC and CQaS, and achieves a comparable good result ($\lambda=0.02$).}
	\label{cleanladybug}
\end{figure*}

\begin{figure*}[htb]
	\centering
	
	\subfigure[SLaT \cite{cai2017three}]{
		\includegraphics[width=1in]{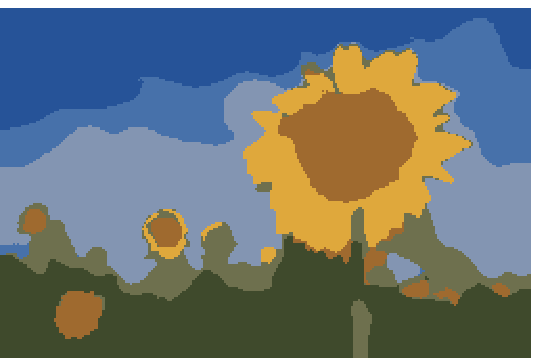}}
	\subfigure[LC \cite{condat2017discrete}]{
		\includegraphics[width=1.in]{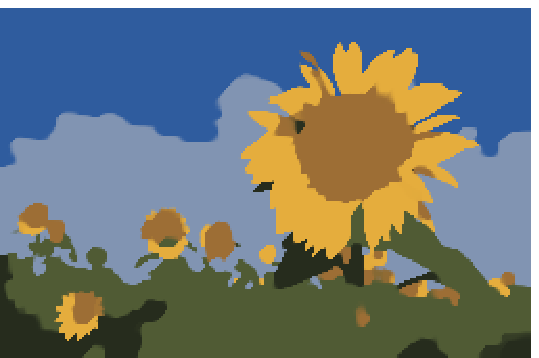}}
	\subfigure[CKC \cite{condat2017convex}]{
		\includegraphics[width=1.in]{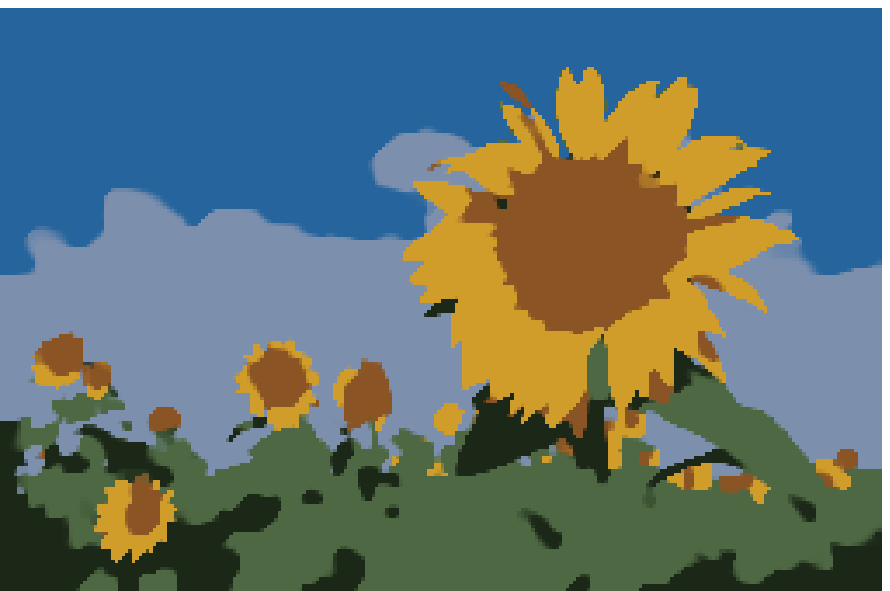}}
	\subfigure[CQaS \cite{condat2017convex}]{
		\includegraphics[width=1.in]{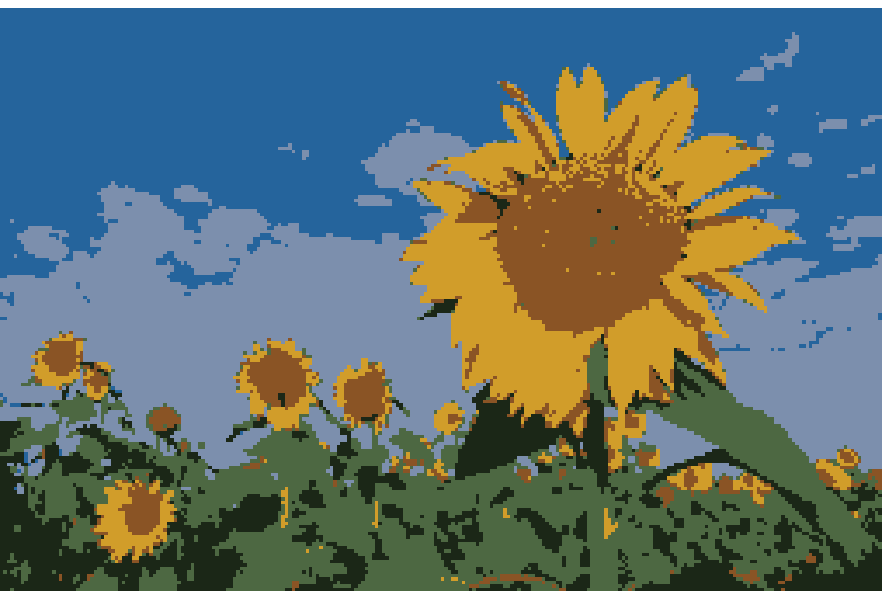}}
	\subfigure[RSSFC \cite{jia2020robust}]{	
		\includegraphics[width=1.in]{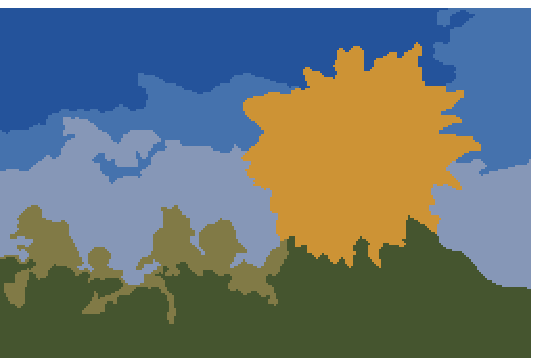}}
	\subfigure[Proposed]{
		\includegraphics[width=1.in]{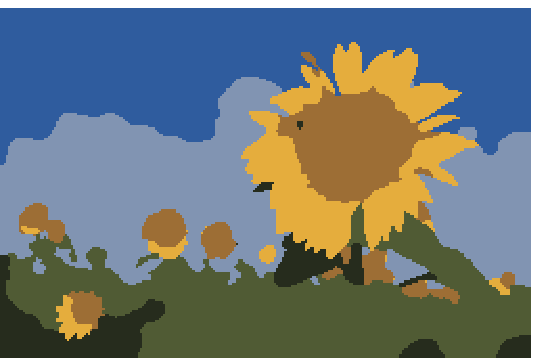}}
	\caption{Segmentation results of Sunflowers in Figure \ref{testimages2}(f). The candidates of palette in this example is shown in Figure \ref{cleansixphases_p}(f). The CQaS (Figure (d)) overcuts the sky; the RSSFC (Figure(e)) undercuts the sunflowers; the SLaT (Figure (a)), LC (Figure (b)), CKC (Figure (c)) and our method (Figure (f)) all have a good result; our method is less complicated than the CKC and achieves a comparable good result ($\lambda=0.09$). }
	\label{cleansunflowers}
\end{figure*}

\section{Experiments}
In this section, to show the superior performance of our proposed method, we compare our method with some methods. The names of these models are the Fuzzy Region Competition (FRC) model \cite{li2010multiphase}, the Dynamic Programming for the Potts (DPP) \cite{storath2014fast}, the Smoothing, Lifting and Thresholding (SLaT) model \cite{cai2017three}, the Potts model for Segmentation (LC) \cite{condat2017discrete}, the Convex K-means Clustering (CKC) model \cite{condat2017convex} and the Robust Self-Sparse Fuzzy Clustering (RSSFC) model \cite{jia2020robust}.
The FRC used fuzzy membership functions to approximate the piecewise constant
Mumford-Shah model. The DPP used the dynamic programming and the ADMM algorithm to solve the Potts model which is non-convex. The SLaT uses a three-stage approach for Segmenting color images (smoothing, lifting and thresholding). The LC proposed a convex approach to K-means without limitation for the segmentation of Parrot. The CKC presented a lifting strategy to solve a piecewise constant Mumford-Shah problem. The RSSFC presented a method of robust self-sparse fuzzy clustering for image segmentation. For all methods in experiments, the scope of the main parameter is provided as follows:
\begin{itemize}
	\item [a.] FRC \cite{li2010multiphase}: the regularization parameter $\lambda \in$ [0.0001, 1];
	\item [b.] DPP \cite{storath2014fast}: the regularization parameter $\lambda \in$ [0.1, 5];
	\item [c.] SLaT \cite{cai2017three}: the regularization parameter $\lambda \in$ [1, 10];
	\item [d.] LC \cite{condat2017discrete}: the regularization parameter $\lambda \in$ [0.001, 10];
	\item [e.] CKC \cite{condat2017convex}: the regularization parameter $\lambda \in$ [1, 500];	
	\item [f.] RSSFC \cite{jia2020robust}: the regularization parameter $\lambda \in$ [0.01, 1];
	\item [g.] Proposed: the regularization parameter $\lambda \in$ [0.0001, 1].				
\end{itemize}

All the experiments were performed under {\sc Matlab} R2020a and Windows 10(x64) on a PC with an Intel Core (TM) i7 10750H CPU at 2.60GHz and 16.0GB of memory.

To quantitative comparison of the segmentation performance, we use one objective indicator, \textit{i.e.,} the segmentation accuracy (SA) \cite{jia2020robust}. The SA is
computed as follows:
\begin{equation}
SA=\sum_{i=1}^{k} \dfrac{S_{i}\cap G_{i}}{n},
\end{equation}
where the $S_{i}$ represents a segmentation result, the $G_{i}$ denotes the corresponding Ground Truth, the $k$ is the number of clusters and $n$ is the total number of pixels of images. \par

\subsection{ Segmentation of synthetic images}
In this subsection, we show the effectiveness of our method on some simple segmentation tasks. These images have obvious different colors and have a reasonable criterion of the segmentation, \textit{e.g.,} we cut the objects from the background. Meanwhile, we compare the robustness of our method with the SLaT on different noise levels. We use the {\sc Matlab} ``imnoise'' command to add Gaussian noise.\par

As aforementioned, the color set could be determined by the K-means method or human. In examples of Three-phases and Six-phases, we use the above two methods to specify the color set. The runtime comparisons show in Table \ref{Table4} between the color set specified by the K-means method and human (including the time taken in deciding the number of colors). We can learn that the runtime of the color set specified by K-means method faster than human, and two methods can arrow the same segmentation accuracy. Thus, we use the K-means method to choose the color set in the following examples.

\textbf{Example 1: Three-phases segmentation.}
Figure \ref{testimages1}(a) is the three-phase synthetic image consisting of three types of shapes. In Figure \ref{cleansixphases1}, \ref{cleansixphases2} and \ref{cleansixphases3}, we add the Gaussian noise with $\tilde{\mu}=0$, $\sigma^2=0.1$, 0.3 and 0.5 ($\tilde{\mu}$: mean, $\sigma^2$: variance) respectively. With the variance of noise increases, the performance of the FRC, SLaT and RSSFC becomes poorer and poorer. We can see that the SLaT has misclassified colors.
 What's more, our method gets more smooth and clear results than the LC and DPP. We can observe that our method is robust for different levels of noise.


\textbf{Example 2: Six-phases segmentation.}
{ Figure \ref{cleansixphases} shows the segmentation results of the six-phase synthetic image consisted of five overlapping circles with different colors. It is natural for us to cut the five overlapping circles with different colors from the background.
 From the results, we can observe that the FRC, DPP and RSSFC fail for the case of the image with Gaussian noise. The FRC is unable to tell the five colors, and the DPP can not keep the shape of the black and green circle and the RSSFC fails to maintain the edges of all circles. Although the SLaT gets a relatively good result, it misclassifies the color of the area between the two neighbor circles.
   Our method is better than the LC as it not only tells the five different colors (see Figure \ref{testimages1}(d)) but also preserves the shape of the circle. }

%

 \subsection{Segmentation of real-world images}
 In this subsection, we test our method on six real-world images to show that our method can get a comparable segmentation result to the CKC without complex calculation. The choice of six real-world images is according to the comparison methods \cite{li2010multiphase, storath2014fast,cai2017three, condat2017discrete, condat2017convex, jia2020robust}.

\textbf{Example 3: Airplane segmentation.}
{We consider a common and simple task of segmenting an image into two parts: foreground (object) and background regions.  We use the airplane image
	  to show the superior performance of our method on this task. The SLaT, LC and our method cut the plane from the sky (background) and segment the image into two reasonable parts. Although this is a simple example, the DPP Figure \ref{cleanairplane}(b) has a poor performance on this task. The FRC and RSSFC do not clearly cut the airplane out as there is a small area that is not separated into the background.
  }

\textbf{Example 4: Hill segmentation.}
{Figure \ref{cleanhill} shows the segmentation results of the hill. The FRC and RSSFC fail to separate the hills as an entirety, the DPP is somewhat under-segmented that some small features have vanished, and the LC does not cut the hill clearly out. The SLaT and our method get relatively better results.}

\textbf{Example 5: Flowers segmentation.}
{Figure \ref{testimages2}(c) consists of four flowers with red, yellow and green color, see Figure \ref{cleansixphases_p}(c). There is no doubt that we want to cut the flowers with different colors from the sky (background). As we can see from Figure \ref{cleanflowers}, for the image with Gaussian noise, the FRC fails to distinguish the colors. The DPP does not make out the objects in the flowers. 
	The SLaT misclassifies the area around the receptacle of the first and third flowers, see the red ``dots''. The LC produces some yellow ``dots''. The RSSFC misclassifies the second flower. Our method gets a relatively good result. }

 {In the above examples, we find that the FRC, FRC and RSSFC perform badly, the SLaT and LC are much superior to the former methods. In the following examples, we mainly compare our method with the SLaT, LC, RSSFC and CKC for three color images shown in \cite{condat2017convex} including the Parrot, the Ladybug and the Sunflowers. \par
 	
 	Here, we emphasize that the method \cite{condat2017convex} specified the $M$ (279 in \cite{condat2017convex}) candidates which takes its values in only $K$ among the $M$ candidates ($K\ll M$). The $M$ colors are shown in Figure \ref{CKClabels}. Meanwhile, the method \cite{condat2017convex} has shown that it is far superior to a two-step strategy: first estimates the $K$ colors using quantization and then solves the segmentation problem restricted to these $M=K$ colors. For convenience, we call this as Coupling Quantization and Segmentation (CQaS) method \cite{condat2017convex}. Although \cite{condat2017convex} proposed an accelerated algorithm and the segmentation results are good, the computational time is still long, as shown in Table \ref{Table2}.}

  {In the next subsections, we use the examples used in the \cite{condat2017convex} to show that our simple modification could get a comparable result to the CKC without complex calculation. For the sake of simplicity, we set $K=6$ in the following examples.}

\textbf{Example 6: Parrot segmentation.}
{In Figure \ref{cleanparrot}, the CQaS and RSSFC have a poor performance obviously. The LC gets a more smooth result than the former method, but it yields a different color. The SLaT undercuts the image, and the area of the eye has been removed. Although our method yields a wrong area in the background, it gets a relatively good result at the object (parrot). What's more, our method with a suitable color set gets a comparable good result to the CKC.}

\textbf{Example 7: Ladybug segmentation.}
{In this example, Figure \ref{cleansixphases_p}(e) shows the palette used. Figure \ref{cleanladybug} shows the segmentation results of the ladybug. The SLaT, LC, CKC, CQaS and our method all have a good result. Here, we emphasize that our method is less complicated than the CKC and achieves a comparable good result to the CKC.}

\textbf{Example 8: Sunflowers segmentation.}
{The candidate of the palette in this example is shown in Figure \ref{cleansixphases_p}(f). Sunflowers have some flowers and the green area at the down of the image, we just want to detect the flowers and take the green area as an entirety. The CQaS overcuts the sky for the given image, and the RSSFC undercuts the sky for the given image. The SLaT, LC, CKC and our method all have a good result.\par
 Here, we highlight that our method is less complicated than the CKC and achieves a comparable good result to the CKC, as shown in Table \ref{Table2}.
}

\section{Conclusion}
In this paper, we propose a new variational segmentation model based on an improved K-means method. We first give a color set determined by human or the K-means method. Then we use a variational model to select the most appropriate colors by SA for each pixel from the color set to get a finer result. The experiments show the effectiveness and robustness of our proposed method. Compared with some state-of-the-art methods, our method has a well running speed while ensuring higher segmentation accuracy. 
	 Recently, deep learning has attracted wide attention. In future work, we intend to use data-driven deep learning to train the regularization and combine the variational model for image segmentation.

\section*{Acknowledgment}

The research was supported by East China Normal University through startup funding grant number 15901-120215-10671, the Natural Science Foundation of China (Grant No. 61971234, 11501301, 62001167), the ``1311 Talent Plan'' of NUPT, Hunan Provincial Key Laboratory of Mathematical Modeling and Analysis in Engineering (Changsha University of Science \& Technology) and Postgraduate Research \& Practice Innovation Program of Jiangsu Province (Grant No. $\mathrm{SJCX19}\_0267, \mathrm{SJCX20}\_0229$).


\bibliographystyle{IEEEtran}
\bibliography{test}

\end{document}